\documentclass[journal,twoside,web]{LaTeX/ieeecolor}
\usepackage{tmi}
\usepackage[noadjust]{cite}  
\usepackage{mathtools,amssymb,amsfonts}
\usepackage{algorithm}
\usepackage{algpseudocode}
\usepackage{graphicx}
\usepackage{textcomp}
\usepackage{multicol,multirow}
\usepackage{booktabs}  
\usepackage{pifont}
\usepackage{hyperref}
\usepackage{colortbl}
\usepackage{dblfloatfix}

\DeclareMathOperator*{\argmin}{arg\,min}

\usepackage[noadjust]{cite}  

\newcommand{\sign}{$^*$}
\newcommand{\signA}{$^\dagger$}
\newcommand{\ns}{\phantom{\sign}}
\newcommand{\nsA}{\phantom{\signA}}

\algtext*{EndWhile}
\algtext*{EndIf}
\algtext*{EndFor}

\newcommand{\red}[1]{\textcolor{black}{#1}}  

\makeatletter
\newcommand{\subalign}[1]{%
  \vcenter{%
    \Let@ \restore@math@cr \default@tag
    \baselineskip\fontdimen10 \scriptfont\tw@
    \advance\baselineskip\fontdimen12 \scriptfont\tw@
    \lineskip\thr@@\fontdimen8 \scriptfont\thr@@
    \lineskiplimit\lineskip
    \ialign{\hfil$\m@th\scriptstyle##$&$\m@th\scriptstyle{}##$\hfil\crcr
      #1\crcr
    }%
  }%
}
\def\logoname{}
\makeatletter
\def\ps@titlepagestyle{%
  \def\@oddhead{\hfil}
  \def\@evenhead{\hfil}
  \def\@oddfoot{\hfil}
  \def\@evenfoot{\hfil}
}
\makeatother

\def\BibTeX{{\rm B\kern-.05em{\sc i\kern-.025em b}\kern-.08em
    T\kern-.1667em\lower.7ex\hbox{E}\kern-.125emX}}
\begin{document}
\title{Regression is all you need for medical image translation}
\author{Sebastian Rassmann, David Kügler, Christian Ewert, Martin Reuter
\thanks{
This work was supported by the German BMFTR (031L0206), the NIH (R01MH131586, R01MH130899, R01AG064027), the CZI (EOSS5 2022-252594), and the Helmholtz Foundation Model Initiative (Human Radiome Project).
We thank the Rhineland Study for providing data and WMH annotations.
}
\thanks{
All authors are with the German Center for Neurodegenerative Diseases (DZNE), Bonn, Germany (e-mail: sebastian.rassmann@dzne.de, david.kuegler@dzne.de, christian.ewert@dzne.de, martin.reuter@dzne.de). 
Martin Reuter is additionally with the A.A.\ Martinos Center for Biomedical Imaging, Boston, MA, USA and the Department of Radiology, Harvard Medical School, Boston, MA, USA.
}}  

\maketitle

\begin{abstract}
\red{
While Generative Adversarial Nets (GANs) and Diffusion Models (DMs) have achieved impressive results in natural image synthesis, their core strengths -- creativity and realism -- can be detrimental in medical applications, where accuracy and fidelity are paramount. These models instead risk introducing hallucinations and replication of unwanted acquisition noise.
Here, we propose \textit{YODA} (You Only Denoise once – or Average), a 2.5D diffusion-based framework for medical image translation (MIT). Consistent with DM theory, we find that conventional \textit{diffusion} sampling stochastically replicates noise. To mitigate this, we draw and average multiple samples, akin to physical signal averaging. As this effectively approximates the DM’s expected value, we term this Expectation-Approximation (ExpA) sampling.
We additionally propose \emph{regression} sampling \textit{YODA}, which retains the initial DM prediction and omits iterative refinement to produce noise-free images in a single step.
Across five diverse multi-modal datasets -- including multi-contrast brain MRI and pelvic MRI-CT -- we demonstrate that \emph{regression} sampling is not only substantially more efficient but also matches or exceeds image quality of full \textit{diffusion} sampling even with ExpA. Our results reveal that iterative refinement solely enhances perceptual realism without benefiting information translation, which we confirm in relevant downstream tasks. \textit{YODA} outperforms eight state-of-the-art DMs and GANs and challenges the presumed superiority of DMs and GANs over computationally cheap regression models for high-quality MIT. Furthermore, we show that \textit{YODA}-translated images are interchangeable with, or even superior to, physical acquisitions for several medical applications.
}

\end{abstract}


\section{Introduction}
\label{sec:introduction}
The ability to learn complex, multi-modal distributions has made Generative Adversarial Nets (GANs) \cite{goodfellow2014generative, isola2017image} and, more recently, Diffusion Models (DMs) \cite{sohl2015deep, song2020score, ho2020denoising} dominant paradigms in natural image generation. 
These models excel at generating highly realistic textures and visually compelling images \cite{ledig2017srgan, Karras_2020_CVPR, dhariwal2021difbeatsgan, saharia2022image, rombach2022ldm}, also leading to their increasing adoption in medical image generation tasks \cite{kazeminia2020gans, kazerouni2023diffusion}.
In medical image translation (MIT), the goal is often to generate synthetic but complementary target modalities from acquired source scans, e.g., to retrospectively augment datasets or accelerate imaging protocols by removing redundancy to reduce the acquisition time.
A typical example in brain MRI is the acquisition of T2 fluid-attenuated inversion recovery (FLAIR), widely used to
detect white matter hyperintensities (WMHs), a biomarker for various neurological conditions \cite{wardlaw2013neuroimaging,lohner2022relation}.  However, due to the inherently slow inversion recovery, FLAIR is often acquired at low resolution \cite{kuijf2019standardized, koch2024rsprotocol, menze2014multimodal_brats} or omitted entirely \cite{glasser2016hcp}.
\red{Thus, MIT offers a compelling way to synthesize such costly or missing sequences.}
Consequently, MIT has become an active research area focused on GANs \cite{dar2019image, dalmaz2022resvit, yu2019ea, benzakoun2022syntheticFlairEAGAN, kong2021breaking, zhang2024unifiedTMI, atli2024i2imamba} and more recently DMs \cite{pinaya2023generative,  meng2024multiTMI, syndiff, arslan2024self, kim2024adaptive, Cho2024SliceConsistent_MICCAI2024, Xin2024Crossconditioned_MICCAI2024,jiang2024fastddpm, zhou2024cascaded, chen2025multiview}.
DMs have shown strong performance, often attributed to their explicit likelihood estimation \cite{syndiff, arslan2024self}, iterative refinement \cite{syndiff}, and the non-deterministic sampling process that is suggested to quantify uncertainty \cite{zhou2024cascaded, chung2022mr}.

\begin{figure*}[t]
    \centering
    \includegraphics[width=0.99\linewidth]{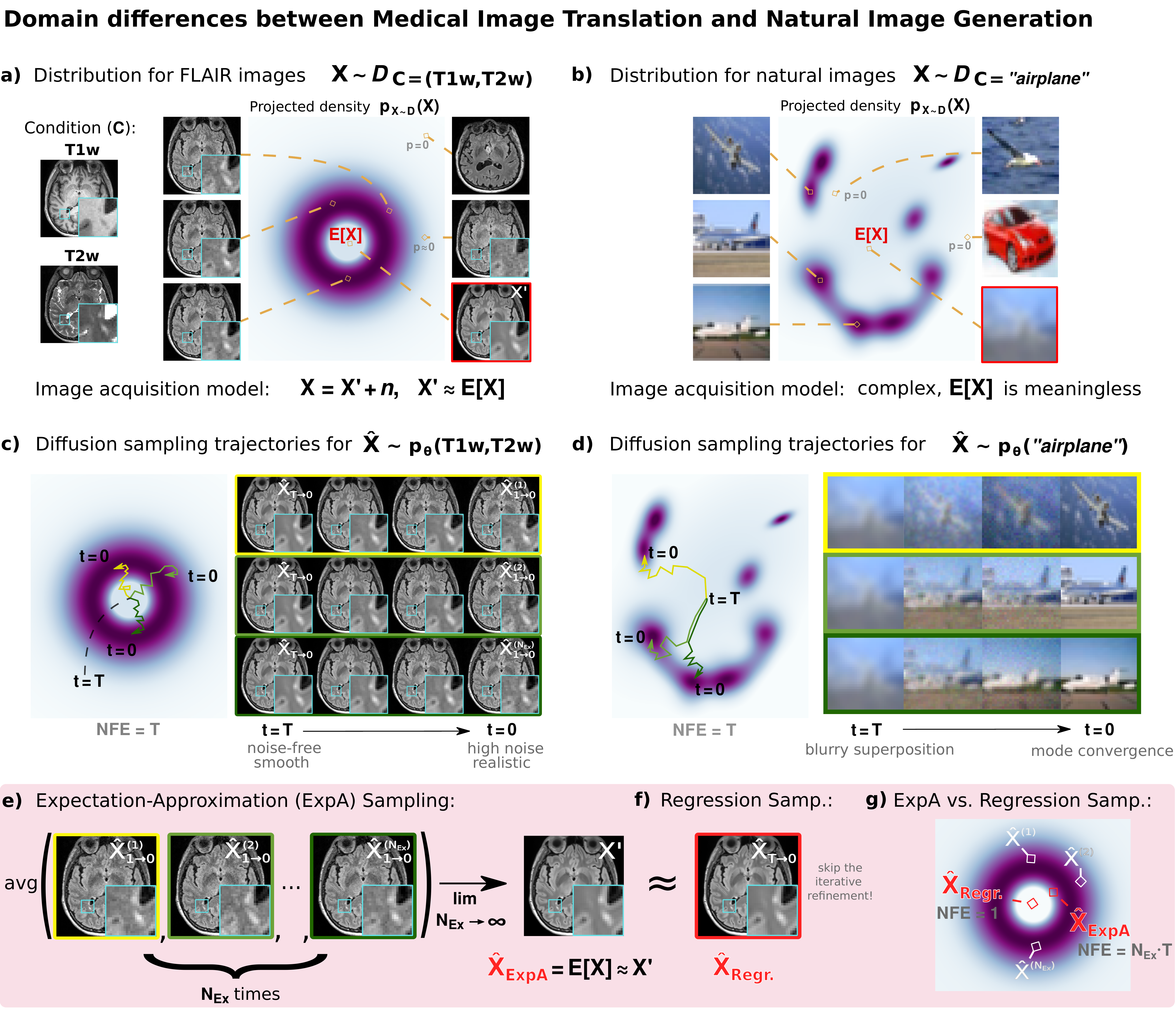}
    \caption{
    \red{
Inherent differences of Medical Image Translation (MIT) and Natural Image Generation (NIG) and the consequence for DDPM sampling. \\
\textbf{(a,b)} Exemplary MIT and NIG image density functions are projected for visualization.
\textbf{(a)} The noisy image acquisition and the strong condition required for faithful MIT (e.g., T1w,T2w $\rightarrow$ FLAIR) creates a uni-modal cluster of acquirable noisy images $X$ centered around the noise-free image $X'$. 
\textbf{(b)} Conversely, the complex image formation and typically weak condition in NIG (e.g., CIFAR-10 \cite{cifar} class conditional) leads to a density landscape with scattered and distinctive modes (compare also Fig.~3 in \cite{ledig2017srgan}). \\
\textbf{(c + d)} Sampling trajectories and exemplary outputs $\hat X_{T\rightarrow0}$ of ideal MIT \textbf{(c)} and NIG \textbf{(d)} denoisers $p_\theta$ (see also \cite{ho2020denoising, karras2022elucidating}).
\textbf{(c)} Note that the generated MIT samples $\hat X_{1\rightarrow 0}^{(j)}$ share the medical information (e.g., lesions) but differ in the noise manifestations (best viewed zoomed in). Thus, image averages are valid ($p>0$) images.
\textbf{(d)} Conversely, NIG sampling branches into distinctive objects of the class (e.g., \textit{"airplane"}) \cite{ho2020denoising, karras2022elucidating}, such that averages are predominantly invalid ($p=0$) images.\\
\textbf{(e)} \textbf{Proposed Expectation-Approximation (ExpA) Sampling (Samp.)}: 
Similar to physical signal averages, uncorrelated noise in samples $\hat X_{1\rightarrow0}^{(j)}, \; j \in [1:N_\text{Ex}]$ drawn from $p_\theta (C)$ enables noise suppression by averaging $X_\text{ExpA}=\text{avg}_j(\hat X_{1\rightarrow 0}^{(j)})$. 
For an ideal $p_\theta(C)$, averaging many samples slowly approaches $\mathbb{E}_{X \sim \mathcal{D}_{C}}\left[X\right]$ (see also the supplementary video). \\
\textbf{(f)} \textbf{Proposed \emph{Regression} Sampling (Regr.)}: 
We simply use the initial minimum-mean-squared-error (MMSE) solution, $\hat X_{T\rightarrow0}$, without further refinement and, thus, noise generation to directly approximate $\mathbb{E}_{X \sim \mathcal{D}_{C}}\left[X\right]$. \\
\textbf{(g)} Whereas ExpA requires full DDPM sampling and exacerbates the intrinsically high number of function evaluations ($N_\text{FE}$) of DDPMs, \emph{regression} sampling\ leverages the single-step MMSE solution requiring only $N_\text{FE}=1$. Note that ExpA and \emph{regression} sampling both approximate the superposition of all valid objects given $C$. While in NIG, this is meaningless \cite{ledig2017srgan,karras2022elucidating}, the strong conditioning and, thus, semantic determinism of MIT make this superposition well-defined with a correspondence to the noise-free image $X'$.
    }}
    \label{fig:overview}
\end{figure*}

Yet, an important distinction must be made between image-to-image (I2I) tasks for natural and medical images \red{(Fig.~\ref{fig:overview}a,b)}.
Whereas natural image generation typically aims to generate realistic and pleasing images \red{from weak conditioning}, medical images are acquired for the accurate depiction of medically relevant information. \red{This demands strong conditioning to preserve underlying anatomical or pathological content, whereas perceptual quality and visual realism become irrelevant, provided that biomarkers and pathologies are faithfully represented}.
Moreover, acquisition noise is a well-known challenge in medical imaging  \cite{holmes1998enhancement, gudbjartsson1995rician, aja2013review}.
Models like GANs and DMs, celebrated for reproducing fine-grained textures and high perceptual quality \cite{saharia2022image, ledig2017srgan}, often risk replicating this noise or hallucinating structures, which in a medical context can cause misleading or even harmful outputs and conclusions. Hallucinations may lead to false findings, while deliberately reproducing noise opposes established efforts of noise suppression during acquisition \cite{holmes1998enhancement, koch2024rsprotocol}. 
\red{Therefore, MIT methods must prioritize accurate information fidelity over visual realism. The sole objective should be to minimize distortion \cite{blau2018perception_distortion_tradeoff}, avoid hallucinations, and suppress noise, ensuring that the generated images remain diagnostically reliable and suitable for downstream analysis.} \\
In contrast to GANs and DMs, Regression Models (RMs) are trained with pixel-wise regression losses (e.g., minimum mean-squared-error, MMSE, models) \red{\cite{jain2008natural, dong2014learning, burger2012image, blau2018perception_distortion_tradeoff}} and approximate the expected value of the data distribution rather than stochastically sampling from a learned probability distribution like DMs and GANs \red{(Fig.~\ref{fig:overview}a)}. Thus, RMs do not replicate non-deterministic image features such as noise \cite{lehtinen2018noise2noise,ledig2017srgan}. 
However, their outputs often appear overly smooth \cite{blau2018perception_distortion_tradeoff, osman2022deep, saharia2022image, saharia2022palette, watson2023novel, isola2017image, ledig2017srgan}, which may introduce a visual domain shift from acquired images and has been suggested impact clinical utility \cite{chung2022mr}.
\red{
Also, using synthetic scintigraphy data, Li et al. \cite{reviewer_paper_synthetic_noise} found that noise removal can improve measured image quality in terms of SSIM and PSNR, while still destroying information. Thus, assessing the actual contained medical information is critical when evaluating MIT results. 
}  

Here, we present a systematic comparison of RMs and DMs for MIT. To this end, we propose \textit{YODA} (You Only Denoise Once - or Average), a novel 2.5D diffusion-based framework operating in uncompressed pixel space and designed as a general-purpose solution for volumetric medical I2I tasks. While \textit{YODA} is trained as a DM, it also supports \emph{regression} sampling by directly using the initial prediction without iterative refinement -- \red{enabling a noise-free, single-step alternative to traditional \textit{diffusion} sampling.}
To systematically compare traditional \textit{diffusion} sampling, which stochastically replicates noise, and noise-free single-step \emph{regression} sampling, we introduce Expectation-Approximation (ExpA) sampling \red{(Fig.~\ref{fig:overview}e)}.
Inspired by signal averaging in physical MRI acquisition, ExpA draws and averages multiple \emph{diffusion} samples to suppress stochastic noise, effectively approximating the model's expected value. \red{Hence, ExpA de-biases the quality assessment of realistic but inherently noisy \textit{diffusion}-generated images compared to smooth \emph{regression} images, providing a principled method to navigate the perception-distortion tradeoff \cite{blau2018perception_distortion_tradeoff}.
} \\
To rigorously evaluate the medical information in generated images, we extend standard assessment protocols by incorporating medically meaningful downstream tasks.
In extensive experiments across five large-scale datasets -- the Rhineland Study (RS) \cite{koch2024rsprotocol} with external evaluation on the \textit{MindBodyBrain} study (MBB) \cite{babayan2019mind}, BraTS \cite{menze2014multimodal_brats, baid2021rsna_brats, bakas2017advancing} and IXI for multi-contrast brain MRI, as well as the Gold Atlas comprising pelvic MRI and CT \cite{nyholm2018gold_atlas} -- we find no systematic advantage of DMs over RMs in translating relevant medical information.
\red{Notably, our results reveal that superior perceptual realism of DMs is largely driven by their replication of acquisition noise, rather than improvements in medically-relevant accuracy. }
This suggests that, in the context of MIT, DMs incur significant computational costs merely to simulate noise. However, noise-free outputs which are preferred for downstream analysis, can be obtained more efficiently through \emph{regression} sampling \red{(Fig.~\ref{fig:overview}f,g)}. 
These findings also challenge the use of DMs for uncertainty estimation, as the observed sampling variance may reflect noise replication rather than true epistemic uncertainty. \\
Building on insights, we demonstrate the clear superiority of \emph{regression}-sampled \textit{YODA} over several strong GAN \red{\cite{dalmaz2022resvit, yu2019ea, atli2024i2imamba}} and DM \red{\cite{syndiff, kim2024adaptive, arslan2024self, Cho2024SliceConsistent_MICCAI2024, chen2025multiview}}  baselines. 
We further show that \textit{YODA}-generated synthetic images are largely interchangeable with real acquisitions in downstream tasks and that the learned modality translation generalizes well to external datasets.

\subsection*{Contributions}
Our main contributions summarize as follows:
\begin{itemize}
    \item We establish \textit{YODA} as a novel 2.5D diffusion framework for medical image translation (MIT).
    \textit{YODA} leverages the volumetric nature of MRI to foster 3D spatial coherence while preserving the computational tractability of uncompressed DMs.
    \item \red{Based on the observed reliability and quality of the initial regression-like DM prediction we introduce \textit{regression} sampling. We show that this achieves the best image quality and performance in downstream applications of all competing methods.}
    \item \red{We propose Expectation-Approximation (ExpA) sampling, a principled method inspired by physical signal averaging. ExpA compensates for noise replication in diffusion sampling enabling a systematic comparisons between diffusion and regression paradigms and revealing their practical equivalence in MIT tasks.}
    \item We demonstrate the superiority of \textit{YODA} over several strong GAN and DM baselines across five diverse MRI and CT datasets. We show that \textit{YODA}-generated images are suitable for downstream medical applications and that \textit{YODA}'s performance generalizes to unseen data sources. 
\end{itemize}


We publish the models at: \href{https://github.com/Deep-MI/YODA}{github.com/Deep-MI/YODA}.

\section{Related work}
Approaches based on conditional GANs (cGANs) \cite{goodfellow2014generative, isola2017image} have long dominated MIT \cite{dar2019image, yu2019ea, kong2021breaking, dalmaz2022resvit, zhang2024unifiedTMI, atli2024i2imamba}.
The initial work on 2D cGANs~\cite{dar2019image} was subsequently extended by ResViT~\cite{dalmaz2022resvit}, \red{I2I-Mamba~\cite{atli2024i2imamba},} and Ea-GANs~\cite{yu2019ea}.
ResViT introduces residual transformer blocks for the generator \cite{dalmaz2022resvit}, 
whereas \red{I2I-Mamba introduces novel dual-domain Mamba blocks with spiral-scan state-space-modeling trajectories \cite{atli2024i2imamba}.} 
Ea-GANs conduct MIT in 3D adding edge information to the GAN loss \cite{yu2019ea} and have shown promising results for deriving FLAIR from diffusion-weighted images 
\cite{benzakoun2022syntheticFlairEAGAN}. 
\red{
To adapt GAN training on degraded images, AmbientGANs explicitly degrade the generated images to learn the distribution of clean images \cite{bora2018ambientgan}. 
While recent work has explored the application of Ambient models to MIT on noisy images \cite{another_reviewer_gan_paper,reviewer_paper_gan,reviewer_paper_dm}, the successful application on real-world images hinges on precisely modeling the noise distribution by including complex artifacts due to accelerated acquisition or motion \cite{aja2013review}.} \\
Recently, DMs \cite{song2020score, ho2020denoising, sohl2015deep} were established as state-of-the-art for various natural image generation tasks \cite{rombach2022ldm, saharia2022image, dhariwal2021difbeatsgan, saharia2022palette}.
DMs operate through many (typically $T \geq 1000$) latent images with \red{increasing or decreasing levels} of noise.
While the forward diffusion process gradually adds noise, the generative backward process iteratively recovers the images using a trained neural network.
Thus, the trained network can generate realistic images from pure noise.
Yet, DM's iterative nature demands a high number of function evaluations (\(N_\text{FE}\)). The resulting high computational costs are exacerbated by the 3D nature of most medical image modalities.
Still, DMs were widely adopted for MIT tasks \cite{syndiff, Xin2024Crossconditioned_MICCAI2024, Cho2024SliceConsistent_MICCAI2024, arslan2024self, meng2024multiTMI, jiang2024fastddpm, zhou2024cascaded, kim2024adaptive, pinaya2023generative, chen2025multiview}. 
The pioneering work of Özbey et al.\ proposes \textit{SynDiff}, an adversarial DM, for unpaired MIT \cite{syndiff}. The authors added GAN losses to compensate for larger denoising steps, reducing the required \(N_\text{FE}\). To a similar effect, \textit{FastDDPM} introduces an accelerated sampling scheme \cite{jiang2024fastddpm}. 
However, we find that accelerated DM sampling reduces high-frequency details (see also the figures in \cite{jiang2024fastddpm}). As we show in Sec.~\ref{sec:noise_theory} and \ref{sec:perc_dist_trade}, this effect biases the reported metrics for image quality, as perceptual quality is traded for reduced distortion \cite{blau2018perception_distortion_tradeoff}. \\
A variant of DMs for I2I tasks is diffusion bridges (DBs) \cite{liu20232i2sb, li2023bbdm}. 
DBs use the conditioning images rather than noise as a prior for the generative diffusion process, which can result in a more direct sampling path \cite{liu20232i2sb, li2023bbdm}. DBs have therefore gained popularity for MIT \cite{arslan2024self, zhou2024cascaded, Cho2024SliceConsistent_MICCAI2024, chung2022mr}. 
An example is the adversarial DM \textit{SelfRDB} \cite{arslan2024self} which introduced a self-consistency step during sampling, i.e.\ editing the image until convergence to improve generation accuracy. CMDM~\cite{zhou2024cascaded} uses a dedicated GAN to generate the DB's prior and additionally employs a multi-path shortcut diffusion strategy, i.e.\ multiple predictions of the DM are averaged to increase translation accuracy and estimate the prediction uncertainty \cite{zhou2024cascaded}. \\
\red{
Several approaches address slice inconsistencies of DMs:
TPDM \cite{lee2023improving} introduced an auxiliary DM perpendicular to the main slicing direction.
Choo et al.~\cite{Cho2024SliceConsistent_MICCAI2024} use a 2D slab-based DB and novel inter-slice trajectory alignment that averages several overlapping predictions. The concurrent work of Chen et al.~\cite{chen2025multiview} proposed MADM which extended slice averaging to several perpendicular denoisers and uses a prior from a 3D GAN to accelerate DM sampling. 
MADM showed promising results for low-dose PET reconstruction \cite{chen2025multiview}.
\\
}
Whereas uncompressed 3D DMs are currently impractical for full-resolution image synthesis due to hardware constraints \cite{pan2023cycle}, a feasible 3D design choice is latent DMs (LDMs) \cite{rombach2022ldm, pinaya2022brain}. Here, DMs operate in the latent space of separately pre-trained autoencoders to reduce the memory footprints. Hence, full-resolution 3D LDMs are realizable on current hardware \cite{kim2024adaptive, jiang2023cola, pinaya2023generative}, with the notable example of ALDM~\cite{kim2024adaptive} that combines the DMs with latent SPADE blocks and showed promising results for MR contrast translation.

Despite findings that simple RMs can effectively learn expectation values of images even if trained solely on noisy samples \cite{lehtinen2018noise2noise}, RMs were rarely considered for MIT. The synthesis results of RMs lacked even anatomically well-defined details such as the gray-white matter (GM/WM) contrast \cite{osman2022deep} or were only applied to small, private datasets, and relatively simple tasks \cite{wei2019fluid}. \\
For I2I tasks on natural images, the evaluation of several early DMs considered training the neural denoiser as RMs baselines \cite{saharia2022image, saharia2022palette, watson2023novel}. This revealed that DMs trade distortion (SSIM and PSNR) for perceptual quality (e.g., Fréchet inception distance, FID, \cite{fid} or human preference \cite{saharia2022palette, saharia2022image}) owing e.g., to the ability to hallucinate high-frequency details \cite{saharia2022image, saharia2022palette, watson2023novel}. 
Whereas ablations of the diffusion process were conducted for ALDM~\cite{kim2024adaptive}, using the neural backbone as a simple RM baseline is commonly omitted \cite{syndiff, Xin2024Crossconditioned_MICCAI2024, Cho2024SliceConsistent_MICCAI2024, arslan2024self, meng2024multiTMI, jiang2024fastddpm, zhou2024cascaded} \red{\cite{reviewer_paper_dm, chen2025multiview, lee2023improving}}.
\red{
An exception are the results of Lyu and Wang \cite{lyu2022conversion} reporting a small benefit of employing DMs of RMs for T2w-to-CT translation using the Gold Atlas dataset.
}
The recent work on medical image denoising showed that single-shot regression sampling of DMs reduces image distortion in comparison to native DM sampling \cite{pfaff2024no}.
\section{Methods}

\subsection{Acquisition noise in quality evaluation} \label{sec:noise_theory}
To illustrate the bias of image metrics for evaluating MIT with noisy images, we analyze its theoretical implications. Modeling MR acquisition as a complex-valued probabilistic process \cite{gudbjartsson1995rician, aja2013review}, the MR magnitude image \(X\) is the combination of a hypothetical, ``true'' image \(X'\) and the complex-valued (primarily thermal acquisition) noise \(n_\text{Re} + i \cdot n_\text{Im}\):
\begin{align} \label{eq:mri_formation}
    X = |X' + n_\text{Re} + i \cdot n_\text{Im} |, \quad  n_\text{Re},n_\text{Im} \sim \mathcal{N}(0, \sigma^2 \mathbf{I}).
\end{align}
Note, however, that this disregards important non-Rician corruptions, e.g., due to accelerated acquisition or motion \cite{aja2013review}. 
Acquiring and averaging \(N_\text{Ex} \gg 1\) independent images suppresses the noise (\(\sigma \propto 1/\sqrt{N_\text{Ex}}\)) and, in theory, allows us to approximate the noise-free image \(X'\) \red{via taking the root-mean-square average (RMS) of multiple images \(X_{i}\)}
\begin{equation}
\begin{aligned} \label{eq:ExpA}
    \lim_{N_\text{Ex} \to \infty} \red{\bar{X}^{(N_\text{Ex})}} &= 
    \lim_{N_\text{Ex} \to \infty} \text{RMS} \left ( X^{(1)}, \; \dots \;, X^{(N_\text{Ex})} \right ) \\
    &=  \lim_{N_\text{Ex} \to \infty} \sqrt{\frac{1}{N_\text{Ex}}  \sum \nolimits_{j=1}^{N_\text{Ex}} 
    X_j^2} \\
    &=  \sqrt{ X'^2 + 2\sigma^2}   = X' \cdot  \sqrt{ 1 + 2{ \cdot (\sigma}/X')^2} \\
    &=  X' \cdot \sqrt{1 + 2 / \text{SNR}^2}  \approx X' ,
\end{aligned}
\end{equation}
\red{where SNR denotes the signal-to-noise ratio}. Assuming a sufficient $\text{SNR} \geq 2$, we simplify the Rician to Gaussian noise \cite{gudbjartsson1995rician, aja2013review}, i.e.\ \(X \sim \mathcal{N}(X', \sigma^2 \mathbf{I}) \).\\
When training GANs and DMs with acquired images \(X \sim \mathcal{X}\), sampling the DM \(p_\theta\) results in ``noisy'' images, as they aim to maximize the likelihood of a generated image \(\hat X \sim p_\theta\) w.r.t.\ \(\mathcal{X}\), i.e.\ produce realistic images. Yet, noise replication influences the assessed image quality: Modeling the generated images \(\hat{X}\) analogously to the acquisition as \(\hat{X}_{\hat{\sigma}} \sim \mathcal{N}(\hat X', \hat{\sigma} ^2 \mathbf{I})\) the MSE decomposes to:
\red{
\begin{equation} \begin{aligned} \label{eq:noiseMSE}
\underset{\substack{\hat{X}_{\hat{\sigma}} \sim p_\theta \\ X \sim \mathcal{X}}}{\text{MSE}} \left(\hat{X}_{\hat{\sigma}}, \, X \right)   
    &= \underset{\substack{
    \hat n \sim \mathcal{N}(0, \hat \sigma^2 \mathbf{I}) \\
    n \sim \mathcal{N}(0, \sigma^2 \mathbf{I}) \\
    }}{\text{MSE}} \left(\hat{X}' + \hat{n}, \, X' + n \right)  
    \\
    &= \mathbb{E} \left [ \left( \hat{X'} - X' \right)^\top \left (\hat{X'} - X' \right ) \right ] \\ 
    & + \mathbb{E} \left[ n^\top n \right] + \mathbb{E} \left[ \hat{n}^\top \hat{n} \right] - 2 \mathbb{E} \left[ n^\top \hat{n} \right] \\ 
    &= \text{MSE} \left( \hat{X'}, X' \right) + \sigma^2 + \hat{\sigma}^2 - 0,
\end{aligned} \end{equation}
where \(\hat n\) denotes deliberately generated noise. 
The measurement noise \(n\) is random by definition and, therefore, unrecoverable, and not correlated with \(\hat n\), which results in \(\mathbb{E} [n^T \hat n] = 0 \).}
Thus, the MSE and the derived PSNR penalize noise creation, favoring solutions with small \(\hat \sigma\). Similar observations can be made for the SSIM. Conversely, reduced \(\hat{\sigma}\) is unrealistic and, therefore, impairs the perceptual quality. 
Hence, the noise amplitude \(\hat{\sigma}\) determines the well-described trade-off between perceptual quality and image distortion, i.e.\ image realism vs. faithful and accurate generation of synthetic images \cite{blau2018perception_distortion_tradeoff}.

\subsection{YODA}
\textit{YODA} is based on the foundational design of DDPMs \cite{ho2020denoising}. The forward process uses the noise schedule \( \beta_t \) for \( t \in [1, T] \):
\begin{align} \label{eq:forward_step}
    q(X_t | X_{t-1}) \sim \mathcal{N} \left(\sqrt{1-\beta_t} X_{t-1}, \beta_t \mathbf{I}\right).
\end{align}  
With \( \Bar{\alpha}_t = \prod_{s=1}^t \left(1 - \beta_s\right) \), this results in latent images
\begin{align} \label{eq:difforward_coeff}
    X_t = \sqrt{\Bar{\alpha}_t} X_0 + \sqrt{1 - \Bar{\alpha}_t} \varepsilon, \quad \varepsilon \sim \mathcal{N}(0, \mathbf{I}) .
\end{align}  
Conversely, the generative backward process follows 
\begin{align}\label{eq:backward_step}
    q(X_{t-1} | X_t) = \int_{X_0} q(X_{t-1} | X_t, X_0) q(X_0 | X_t) \, dX_0 
\end{align}
with \( q(X_{t-1} | X_t, X_0) \) from the forward process \eqref{eq:forward_step} and \( q(X_0 | X_t) \) incorporates knowledge of the data distribution \(\mathcal{X}\). In practice, the image \(\hat X_0\) is approximated iteratively by predicting the acquired image (\(\hat{X}_{t \rightarrow 0} \approx X_0\)) at each time step \(t\) using a trained neural network \(p_\theta\) as
\begin{align} 
    p_\theta(\hat X_{t-1} | \hat X_t) = \sqrt{\Bar{\alpha}_t} \hat{X}_{t \rightarrow 0} + \sqrt{1 - \Bar{\alpha}_t} \varepsilon_t \rightarrow \hat X_{t-1} \label{eq:diffbackward}
\end{align}  
with an initial starting value \( \hat{X}_T \sim \mathcal{N}(0, \mathbf{I}) \). \\
Whereas DMs usually predict either the noise-free image \( X_0 \) or the noise \( \varepsilon_t \), \textit{YODA} predicts the velocity
\(v_t = \sqrt{\Bar{\alpha}_t} \varepsilon - \sqrt{1 - \Bar{\alpha}_t} X_0 \) \cite{v_pred}.
\red{Denoting the source conditioning images as $C$, the joint distribution of source and target images \( \mathcal{X}_C \), and the uniform distribution of time steps as \( \mathcal{U}_T \)}, the training objective for \(v_\theta\) is the empirical risk minimization of the velocity differences
\begin{align} \label{eq:diff_obj_func}
    \min_{\theta}  \;
    \mathbb{E}_{
    (X_0, C) \sim \mathcal{X}_C, \,
    \varepsilon \sim \mathcal{N}(0, \mathbf{I}), \,
    t \sim \mathcal{U}_{T} 
    } 
    \left[
    L_2(v_\theta(X_t, t, C), v_t)
    \right].
\end{align}
This objective can be derived from the evidence lower bound of the probabilistic model \cite{ho2020denoising}. 
Training and native DDPM sampling are described in Alg.~\ref{algo:train} and \ref{algo:native_sample}: \algrenewcommand\algorithmicensure{\textbf{Optional:}}

\begin{algorithm} 
\caption{Training \textit{YODA}} \label{algo:train}
\begin{algorithmic}[1] \ttfamily \footnotesize{
\For{$(X, C) \sim \mathcal{X}_C$, $\varepsilon \sim \mathcal{N}(0, \mathbf{I})$, $t \sim \mathcal{U}_{T}$} 
    \State $X_t = \sqrt{\bar{\alpha}_t}X + \sqrt{1-\bar{\alpha}_t}\varepsilon$ 
    \hfill \# forward diffusion
    \State $v_t = \sqrt{\bar{\alpha}_t}\varepsilon - \sqrt{1-\bar{\alpha}_t}X$ 
    \hfill \# $v$-prediction target
    \State Gradient descent step on: \( \nabla_\theta \: L_2 \left (v_\theta (X_t, \, t, \, C),  v_t \right ) \)    
\EndFor }
\end{algorithmic}
\end{algorithm}
\vspace{-0.5cm}

\begin{algorithm} 
\caption{Native DDPM sampling}  
\label{algo:native_sample}
\begin{algorithmic}[1] \ttfamily \footnotesize
\Require $C$: conditioning images 
    \State $X_T = \varepsilon_T \sim \mathcal{N}(0, \mathbf{I})$
    \For{$t$ in  $\left ( T, \; T-1, \; \ldots \;, \; 1 \right )$}  \hfill \# $N_\text{FE}=T$
        \State $\hat{X}_{t\rightarrow 0}  = \sqrt{\bar{\alpha}_t} \, \hat{X}_t - \sqrt{1 - \bar{\alpha}_t} \, v_\theta (\hat{X}_t, \, t, \, C)$
        \If{$t > 1$}   
            \State $\varepsilon_{t-1} \sim \mathcal{N}(0, \mathbf{I})$
            \State $\hat{X}_{t-1} = \sqrt{\bar{\alpha}_{t-1}} \, \hat{X}_{t\rightarrow 0} + \sqrt{1-\bar{\alpha}_{t-1}} \, \varepsilon_{t-1}$
        \EndIf
    \EndFor
\State \textbf{return} $\hat{X}_{1\rightarrow 0}$

\end{algorithmic} 
\end{algorithm}

\red{
Based on the introduced concepts and notation, we describe the extension of \textit{YODA} sampling over default DDPMs with additional optional arguments in Alg.~\ref{algo:sample}:
\begin{algorithm} 
\caption{Sampling \textit{YODA}}  
\label{algo:sample}
\begin{algorithmic}[1] \ttfamily \footnotesize
\Require $C$: conditioning images 
\Ensure $N_\text{Ex}=1$, $T_\text{trunc.} = T-1$        \hfill \# default: \textit{diffusion}
\For{$i$ in $\left ( 1, \; \dots \, , \; N_\text{Ex} \right )$} 
    \State $X_T = \varepsilon_T \sim \mathcal{N}(0, \mathbf{I})$
    \For{$t$ in  $\left ( T, \; T_\text{trunc.}, \; \ldots \; , \; 1 \right )$}  \hfill \# $N_\text{FE}=T_\text{Trunc.} + 1$
        \State $\hat{X}_{t\rightarrow 0} = \hat{X}^{(i)} = \sqrt{\bar{\alpha}_t} \, \hat{X}_t - \sqrt{1 - \bar{\alpha}_t} \, v_\theta (\hat{X}_t, \, t, \, C)$
        \If{$t > 1$}   
            \State $\varepsilon_{t-1} \sim \mathcal{N}(0, \mathbf{I})$
            \State $\hat{X}_{t-1} = \sqrt{\bar{\alpha}_{t-1}} \, \hat{X}_{t\rightarrow 0} + \sqrt{1-\bar{\alpha}_{t-1}} \, \varepsilon_{t-1}$
        \EndIf
    \EndFor
\EndFor
\State \textbf{return} $\text{RMS} \left (\hat{X}^{(1)}, \; \dots \, , \; \hat{X}^{(N_\text{Ex})} \right )$
\end{algorithmic} 
\end{algorithm}

} \\
The modifications are detailed in the following:

\subsubsection{2.5D diffusion} \label{sec:25D_diff} \begin{figure}
    \centering
    \includegraphics[width=0.99\linewidth]{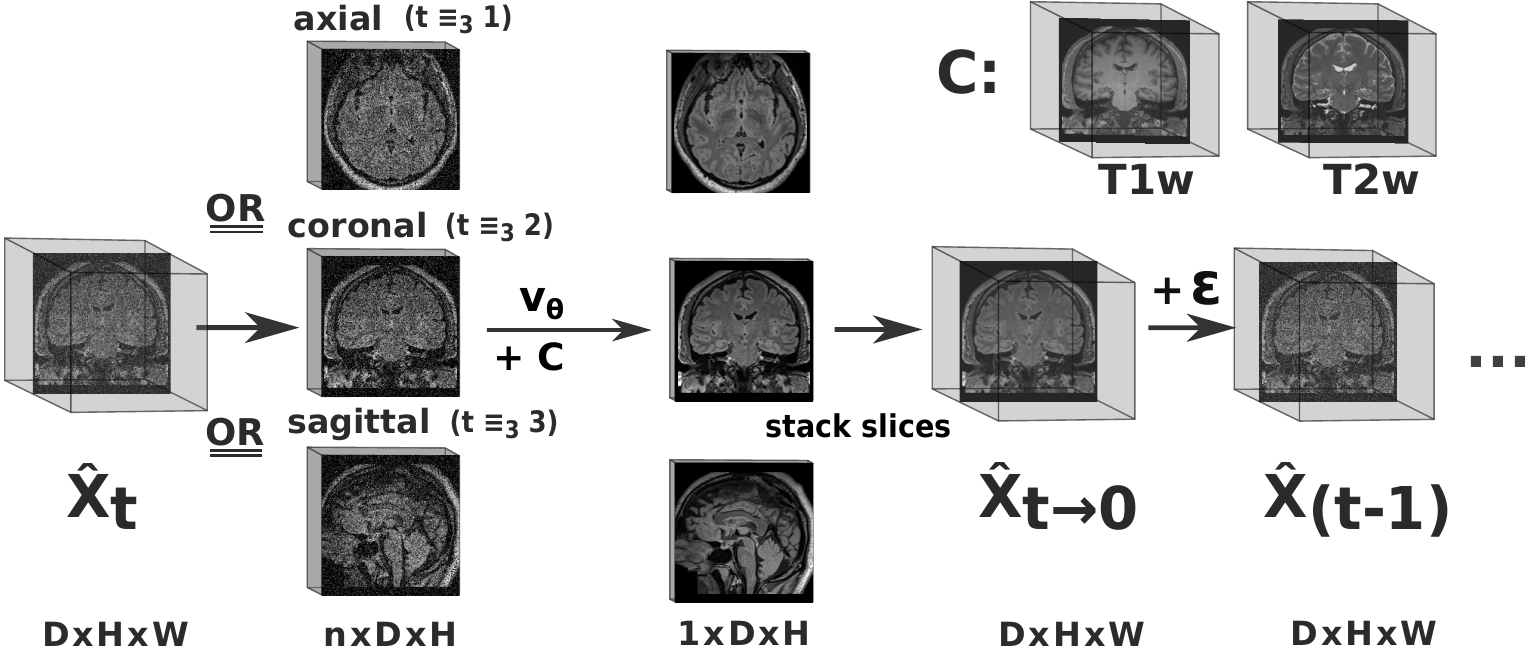}
    \caption{\red{
    2.5D \emph{diffusion} sampling of \textit{YODA}: In each time step $t$, the latent diffusion volume $X_t \in \mathbb{R}^{D\times H \times W}$ is processed in slabs of  $n$ consecutive slices (default: $n=5$). 
    The slabs are concatenated with the conditioning $C$, i.e.\ the corresponding slabs of the source images as input the neural denoiser $v_\theta$. The output slices are then stacked as predicted volume $\hat X_{t\rightarrow0}$ to which noise $\varepsilon$ is added to obtain the next latent image $\hat X_{t-1}$.
    Note, that the slicing plane is rotated between axial, coronal, and sagittal. 
    }} \label{fig:25D}
\end{figure} 
Operating the neural denoiser on full-resolution 3D images is unfeasible on current hardware due to memory constraints. Thus, we perform a 2.5D diffusion approach \red{(Fig.~\ref{fig:25D}):
We define the diffusion process in 3D, i.e.\ the latent diffusion images are volumes \(X_t \in \mathbb{R}^{D \times H \times W}\).
The 2.5D denoiser \(v_\theta\) ingests 2D slabs 
\( \{X_{t,1}, \, \dots \, X_{t,D} \}\) with }\red{\(X_{t,i} \in \mathbb{R}^{n \times W \times H}\), as inputs and predicts slices 
$\hat{X}_{t\rightarrow 0,i} \in \mathbb{R}^{1 \times W \times H} $. \(v_\theta\) implicitly stacks these slices to form the predicted noise-free volume 
\(\hat{X}_{t \rightarrow 0} \in \mathbb{R}^{D \times H \times W} \).} \color{black}
Based on this, we include the following 2.5D components:
\paragraph{multi-slice inputs}
\(v_\theta\) predicts a single slice from a slab formed with the two bi-directional adjacent slices (i.e.\ 5 in total) of both, the conditioning \(C\) and the diffusion latent \(\hat X_t\). 
\red{This allows for a harmonization of image appearance and feature creation across slices during the sampling process.}  
\paragraph{orthogonal denoising}
We rotate the slicing views (sagittal, coronal, axial) between denoising steps to improve 3D information transfer and combat slicing artifacts \red{(Fig.~\ref{fig:25D})}. 
\red{For simplicity and to avoid extra training effort, all views are predicted by a single network.}  
\red{Note that \textit{YODA}'s 2.5D sampling scheme does not increase the required $N_\text{FE}$ due to multi-plane averages or correction steps \cite{lee2023improving, chen2025multiview}.}
\subsubsection{Truncated sampling} \label{sec:lazy_sampl} \begin{figure}[b]
    \centering
    \includegraphics[width=0.99\linewidth]{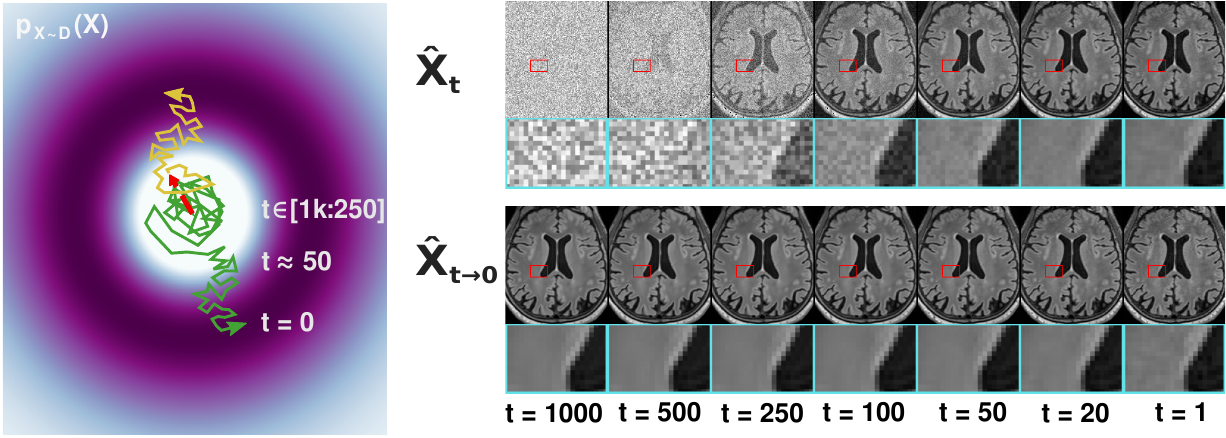}
    \caption{\red{
    In regular DM sampling (green), the diffusion prior \(\hat X _t\) for time steps \(t {\in}[1k{:}250]\) is weak compared to the conditioning \(C\). Thus, the estimate of the noise-free images (\(\hat X_{t\rightarrow0}\), line 3 of Alg.~\ref{algo:native_sample}) remains almost unchanged (best view zoomed-in). We therefore truncate DM sampling skipping the time steps between $t{=}999$ and $250$ (red arrow).
    }} \label{fig:trunc}
\end{figure}

As the initial solution \(\hat{X}_{T\rightarrow 0}\) remains practically unchanged for many late diffusion steps \red{(Fig.~\ref{fig:trunc})}, we truncate DM sampling: we skip the following $\sim 75\%$ low-SNR denoising steps, i.e.\ we only consider $t \in \{T, T/4, \ldots, 1 \}$ \red{by setting $T_\text{Trunc.} = T/4$ in the modified Alg.~\ref{algo:sample}}.
Note that in contrast to CMDM~\cite{zhou2024cascaded} \red{and MADM~\cite{chen2025multiview}}, we derive the truncation directly from the backward process, re-using the initial DM solution, \(\hat{X}_{T\rightarrow 0}\), as the prior.

\subsubsection{Expectation-approximation (ExpA) sampling}
DM sampling is non-deterministic \red{(illustrated in Fig.~\ref{fig:overview}c,d)} due to adding random noise in the backward process (\(\varepsilon_t\) in \eqref{eq:diffbackward} and Alg.~\ref{algo:native_sample} and ~\ref{algo:sample}). 
Thus, for each run \red{$j \in \{1, \dots, N_\text{Ex} \}$}, 
\red{each drawn sample, i.e.\ last image of the backward process,}
\red{\(\hat{X}^{(j)}_{1\rightarrow0} \sim p_\theta\)} has a different noise manifestation $\hat{n}_i$ \red{(Fig.~\ref{fig:overview}c)}. 
We can thus approximate the DM's expected value \red{using ExpA sampling ($N_\text{Ex}>1$ for Alg.~\ref{algo:sample})} analogously to acquisition averages \eqref{eq:ExpA}: 
\red{
\begin{align}
{\hat{X}}_{\text{ExpA}}^{(N_\text{Ex}) } = \text{RMS}\left (\hat{X}^{(1)}, \, \dots \, , \ \hat{X}^{(N_\text{Ex})} \right ).
\end{align}}
For an ideal DM (i.e.\ \(
\text{KL}\left [p_\theta(C) \, \| \, p(X | C) \right ] = 0
\)), we would obtain 
\begin{align}
        \mathbb{E}_{\hat X \sim p_{\theta}(C)} 
        \!\! \overset{\hat n_i\sim\mathcal{N}}{=}  \!\!
        \lim_{N_\text{Ex} \to \infty} \! {\hat{X}}_{\text{ExpA}}^{(N_\text{Ex}) }   \,\,\overset{\text{KL}=0}{=} 
        \lim_{N_\text{Ex} \to \infty} \! {\bar{X} }^{(N_\text{Ex})}
        \,\, \overset{\text{\eqref{eq:ExpA}}}{\approx} \,\,X'.
\end{align}
\red{
Unlike previous works averaging the predicted noise-free images \(\hat X_{t \rightarrow 0}\) \cite{Cho2024SliceConsistent_MICCAI2024} or diffusion latents $X_{t-1}$ \cite{chen2025multiview} for $t{>}1$, the proposed ExpA sampling only averages the final samples $\hat X_{1\rightarrow 0}$. 
Thus, ExpA sampling decouples (implicit) noise-suppressing averages and 3D coherency mechanisms to sustain the SNR in the backward process such that, assuming an ideal $p_\theta$,
 \(
\text{KL} [ p_\theta( \hat X_{t}, C) \, \| \, q(X_{t-1} |  \, C) ] = 0
\).
Therefore, the proposed ExpA sampling follows physical signal averages as a clear paragon. \\ 
Furthermore, previous works proposing averages of DM samples neither described the relation to physical noise suppression nor the approximation of the expected value of the DM, whereas the divergence of DM sampling was attributed to model uncertainty \cite{zhou2024cascaded, lyu2022conversion}. }




\subsubsection{Efficient approximation of expected values}
Let us now investigate the behavior of DMs for the first denoising iteration, i.e.\ $t = T$, in more detail.
Since $\bar{\alpha}_T \approx 0$, DMs receive pure, uninformative noise $\varepsilon$ as the diffusion prior.
Denoting the joint distribution of noise-free targets and conditions as $\mathcal{X}'_C$, the training objective at $t=T$ simplifies to: 
\begin{align} \label{eq:rm_obj_func}
    \min_{\theta} \,
    \mathbb{E}_{
    (X', C) \sim \mathcal{X}'_C, n \sim \mathcal{N}(0, \sigma^2 \mathbf{I})
    } 
    \left[
    L_2(v_\theta(C), \; X' + n )
    \right].
\end{align}
The theoretical minimum of this objective is assumed \red{at the MMSE solution, i.e.\ \(\hat{X}_{T\rightarrow0} \approx \mathbb{E}[X] \approx X'\) and, thus,} a perfect DM would already predict noise-free images in the first sampling operation \red{(Fig.~\ref{fig:overview}e)}. This provides a tractable alternative to sampling the DM with $N_\text{Ex}{=}\infty$ \red{(Fig.~\ref{fig:overview}f)}. We therefore also considered this initial solution of \textit{YODA} as a sampling method for approximating noise-free images, i.e.\ \(\hat{X}_{T\rightarrow0} \approx X'\).
\red{As this sampling only uses the regressive single-shot abilities of the DM and is expected to have the same image properties as applying dedicated RMs,} we term this \emph{regression} sampling \red{($T_\text{trunc.}=0$ in Alg.~\ref{algo:sample})}. \\
Note that, in general, RMs can be seen as a special case of DMs, where setting $T=1$ during the training only optimizes the single-step \emph{regression} prediction from the source images (with only vacuous noise as additional input). 

\subsubsection{3D coherency in \emph{regression} sampling} 
As $v_\theta$ operates in 2D, \emph{regression} sampling lacks access to 3D information. Thus, to improve the performance of \emph{regression} sampling, we aggregate the predictions of all three views in our final model. \red{We omit view aggregation and use only axial slices for the ablation experiments where not otherwise noted.}
Additionally, 2.5D diffusion homogenizes the appearance of individual slices due to orthogonal denoising (Sec.~\ref{sec:25D_diff}). For a similar effect for \emph{regression} sampling, we propose to correct the individual slices \(\hat X_j)\) that form the volumes via gamma corrections
\begin{align} \label{eq:gamma}
    \Gamma_{\theta_j}(\hat{X}_j) = (1 + a_j) \hat{X}_j^{(1 + \gamma_j)} + c_j,
\end{align}
with a set of parameters \(\theta_j = \{a_j, \gamma_j, c_j \}\) for each slice \(\Hat{X}_j\).
We then optimize the parameters \(\theta_j\) to minimize the intensity differences between adjacent slices \(\hat{X}_j\) and \(\hat{X}_{j+1}\):
\begin{align}
    \argmin_{\{\theta_0, \dots, \theta_{N-1} \}} 
    \frac{1}{N-1} \sum_{j=0}^{N-2} L_2 \left ( 
    \Gamma_{\theta_j}(\hat{X}_j), \Gamma_{\theta_{j+1}}(\hat{X}_{j+1})
     \right )  \label{eq:zsmooth}
\end{align}

\subsection{Datasets and preprocessing}
We obtained a stratified set of 2233 participants of the RS \cite{koch2024rsprotocol}, which were split into 1344 for training, 237 for validation, and 600 for testing. 52 additional participants with manual WMH annotation \cite{lohner2022relation} constituted the WMH test set. MR acquisition was conducted at two sites using the same 3T scanner type and protocol \cite{koch2024rsprotocol}. We also tested on 82 MBB participants for whom T1w, T2w, and 1~mm FLAIR images were available \cite{babayan2019mind}.
T1w and T2w were registered (RS: \textit{mri\_coreg}, MBB: \textit{bbregister}) to the FLAIR images and interpolated (cubic) to 1~mm with \textit{FreeSurfer} (v7.4) \cite{fischl2012freesurfer,greve2009bbreg}. \\
The IXI dataset (\href{www.brain-development.org/ixi-dataset}{brain-development.org/ixi}) contains T1w, T2w, and proton-density (PD) images from 577 participants (111 withheld for testing) acquired at three different sites equipped with 1.5 or 3T scanners from two vendors. We registered the T1w and PD to the T2w images with \textit{mri\_coreg}. \\
\red{
For the BraTS 2023 challenge \cite{menze2014multimodal_brats, baid2021rsna_brats, bakas2017advancing}, we assembled a test set of cases unseen to the downstream tool (trained on BraTS 2020) \cite{ isensee2021nnu, isensee2021nnu_general} and with public reference tumor segmentation labels. To this end, we tested on 200 randomly selected training cases of the 2023 challenge that were added \emph{after} 2020. We then trained on the remaining training cases of the 2023 split.
The validations set was used for the ablations, whereas we added the validation data to the training set of the final models.} 
The BraTS dataset includes data from various centers and protocols, and is provided as normalized and skull-stripped images resampled to a 1~mm resolution. We derived full-brain segmentation masks and, for all but the BraTS dataset, performed robust intensity normalization with \textit{FastSurfer} (v2.2) \cite{henschel2020fastsurfer, henschel2022fastsurfervinn}. \\
MRI-to-CT translation was performed in 15 participants (4 for testing) of the Gold Atlas \cite{nyholm2018gold_atlas}, which provided co-registered CT, T1w, and T2w images. We created tissue masks based on intensity thresholds on the T1w images and excluded the upper and lower 5 slices due to decreased quality. \\
We defined the region-of-interest (ROI) for translation as the 3D bounding box of the tissue masks expanded by 5, 10, and 20 (axial, coronal, sagittal) voxels. 

All participants gave informed consent to the respective studies.

\begin{table*}[!b] \centering \begin{scriptsize} 
\caption{Performance for RS-trained brain MRI translation (T1w,T2w $\rightarrow$ FLAIR) for in- and external (MBB) test sets.
\sign \ and \signA \ mark significant ($p<0.05$) differences to regression sampling YODA and to WMH segmentation on acquired images, respectively. \ \red{¹ $\cdot 10^{-3}$, a.u., \; ² smaller (54M), 2D-only version, \;  ³ in a compressed latent space}
} \label{tab:rs_metrics}

\begin{tabular}{
l @{\hskip 0.2cm} l @{\hskip 0.01cm} c @{\hskip 0.3cm} 
c @{\hskip 0.13cm} c @{\hskip 0.13cm} r @{\hskip 0.5cm}  
c @{\hskip 0.10cm} c @{\hskip 0.2cm}  
| @{\hskip 0.3cm}   
c @{\hskip 0.1cm} c @{\hskip 0.2cm} c}
\toprule
 & \multicolumn{2}{l}{required} & \multicolumn{3}{c}{test set ($n=600$)} & \multicolumn{2}{c}{MBB set ($n=82$)} & \multicolumn{3}{c}{WMH test set ($n=52$, manual reference)} \\
 & time & \red{(NFE)} & SSIM ($\%, \uparrow$) & PSNR ($\text{dB}, \uparrow$) & \red{FID¹ (\(\downarrow\))} & SSIM ($\%, \uparrow$) & PSNR ($\text{dB}, \uparrow$) & Dice \cite{shiva_wmh} ($\%, \uparrow$) & ALVR \cite{shiva_wmh} ($\downarrow$) & CNR ($\%, \uparrow$) \\ \midrule
\textit{YODA} (\emph{regr.}) & 20 s & (3) & \textbf{97.31 ± 0.90}\ns & 33.66 ± 1.95\ns & 16.26 & 93.25 ± 1.14\ns & 26.75 ± 1.19\ns & \textbf{58.72 ± 18.51}\ns\ns & \textbf{0.30 ± 0.25}\signA\ns & 3.54 ± 0.51\ns\ns \\
\textit{YODA} (\(N_\text{Ex}{=}10\)) & 2.5 h & (2.5k) & 97.19 ± 0.93\sign & \textbf{33.79 ± 1.82}\sign & 12.59 & \textbf{93.27 ± 1.16}\ns & 26.76 ± 1.22\ns & 58.48 ± 18.70\sign\signA & \textbf{0.30 ± 0.23}\signA\ns & 3.48 ± 0.49\sign\ns \\
\textit{YODA} (\(N_\text{Ex}{=}4\)) & 1 h & (1k) & 96.99 ± 0.97\sign & 33.57 ± 1.76\sign & 8.69 & 93.14 ± 1.16\sign & 26.70 ± 1.20\sign & 58.21 ± 18.62\sign\signA & 0.31 ± 0.24\signA\ns & 3.42 ± 0.48\sign\signA \\
\textit{YODA} (\emph{diff.}) & 15 m & (251) & 96.07 ± 1.12\sign & 32.66 ± 1.53\sign & 0.96 & 92.46 ± 1.20\sign & 26.42 ± 1.14\sign & 56.12 ± 19.10\sign\signA & 0.37 ± 0.30\sign\signA & 3.18 ± 0.45\sign\signA \\ 
\midrule
\red{\textit{YODA}² (2D, regr.)} & 3 s & (1)  &  96.80 ± 0.97\sign &  33.23 ± 1.47\sign & 19.68 & 91.86 ± 1.14\sign &  25.70 ± 1.09\sign &  56.69 ± 19.63\sign\signA &  0.34 ± 0.23\ns\ns &  3.34 ± 0.59\sign\signA \\
\midrule
\red{Choo et al.} \cite{Cho2024SliceConsistent_MICCAI2024} & 45 m & (100) & 96.94 ± 0.96\sign &  32.98 ± 2.25\sign & 23.76 & 93.08 ± 1.13\sign &  27.51 ± 1.51\sign  & 56.97 ± 20.44\sign\signA &     0.41 ± 0.39\sign\ns &  \textbf{3.66 ± 0.60}\sign\signA\\
\red{MADM} \cite{chen2025multiview} & 2.2 h & (603) & 96.40 ± 1.08\sign &  32.68 ± 1.76\sign & 16.38 &  93.20 ± 1.36\ns &  \textbf{27.90 ± 1.75}\sign  & 57.72 ± 19.10\sign\signA &    0.33 ± 0.25\signA\ns &     3.43 ± 0.50\sign\ns \\
\textit{SelfRDB} \cite{arslan2024self} & 90 s & (20) & 95.68 ± 1.10\sign & 30.75 ± 1.43\sign & 15.78 & 92.72 ± 1.17\sign & 27.12 ± 1.37\sign & 55.73 ± 21.12\sign\signA & 0.42 ± 0.36\sign\ns & 3.32 ± 0.66\sign\signA \\
\textit{SynDiff} \cite{syndiff} & 10 s & (4) & 96.18 ± 1.06\sign & 32.42 ± 1.57\sign & \textbf{0.69} & 92.76 ± 1.24\sign & 27.32 ± 1.49\sign & 55.65 ± 19.30\sign\signA & 0.45 ± 0.36\sign\ns & 3.27 ± 0.53\sign\signA \\
ALDM \cite{kim2024adaptive} & 60 s & (250)³ & 87.92 ± 1.39\sign & 26.57 ± 2.15\sign & 10.69 & 89.03 ± 1.86\sign & 25.77 ± 1.66\sign & 32.35 ± 22.86\sign\signA & 1.15 ± 0.89\sign\signA & 1.32 ± 0.57\sign\signA \\
\red{I2I-Mamba} \cite{atli2024i2imamba} & 15s & (1) & 95.61 ± 1.11\sign &  31.75 ± 1.44\sign & 1.51 & 92.99 ± 1.16\sign &  27.58 ± 1.43\sign  & 57.07 ± 18.68\sign\signA &     0.35 ± 0.29\sign\ns &  3.11 ± 0.50\sign\signA
\\
ResViT \cite{dalmaz2022resvit} & 10 s & (1) & 94.72 ± 1.13\sign & 30.52 ± 1.33\sign & 5.45 & 90.15 ± 1.49\sign & 25.66 ± 1.60\sign & 54.04 ± 19.55\sign\signA & 0.50 ± 0.41\sign\ns & 2.96 ± 0.52\sign\signA \\
Ea-GAN \cite{yu2019ea} & 5 s & (1) & 95.76 ± 0.97\sign & 31.45 ± 1.46\sign & 11.09 & 92.14 ± 1.08\sign & 26.79 ± 1.32\ns & 42.86 ± 20.29\sign\signA & 0.92 ± 0.66\sign\signA & 2.54 ± 0.54\sign\signA \\ 
\midrule
Acquired & 4.5 m & \ - & 100 & - & - & 100 & - & 59.44 ± 17.39\ns\ns & 0.43 ± 0.34\sign\ns & 3.50 ± 0.39\ns\ns \\ \bottomrule
\end{tabular} \end{scriptsize}
\end{table*}

\subsection{Implementation details}  \label{sec:implementation}
YODA is parametrized by a U-Net \cite{pinaya2023generative, rombach2022ldm} with 5 ResNet blocks (128/128/256/256/512 channels) in the en- and decoder, group normalization (group size 32), SiLU activation functions, and residual self-attention layers with time-step embedding (0/0/1/1/2 heads), resulting in 53M parameters. For our final RS model (Tab.~\ref{tab:rs_metrics}), we increase the size to 214M parameters. Source image conditioning is realized via channel-wise concatenation \cite{saharia2022image}. 
We use the linear \cite{ho2020denoising} (default) and cosine noise schedules \cite{nichol2021improved} with $T=1000$. \\
We train \textit{YODA} with an ADAM optimizer with a learning rate (LR) of \(10 ^{-4} \) (chosen from \(  \{ 10 ^{-3}, 10 ^{-4}, 10 ^{-5}  \}  \)), a batch size of 96 (Gold Atlas: 24), and track the parameters' exponential moving average with a decay factor of 0.999 (from \( \{0, 0.99, 0.999, 0.9999\} \)). 
To ensure a balanced training effort, we randomly sample the slicing plane and then a slice within the ROI. In total, we sample each volume 20k times, i.e.\ each slice, on average, 125 times.\\
In line with our baseline methods, 
we omit skull-stripping for IXI. For training \textit{YODA} on the RS, we reduce the weight of voxels outside the brain mask to a small \(\varepsilon = 0.01 \), which retains the skull for downstream uses while focusing on actual brain translation. 
As this is not possible for adversarial losses, we skull-strip the images for a fair comparison. \\
We perform slice homogenization following \eqref{eq:zsmooth} for BraTS \emph{regression} sampling by optimizing \(\theta_i\) with stochastic gradient descent (\( \text{LR} = 10\)). 
We exclude background voxels and add a small MSE penalty (\(\varepsilon = 0.01\)) for \(\theta_i\) to the loss. The optimization converges within $2000$ steps ( \({<}10\) seconds).

\subsection{Competing methods} 
We selected baseline methods based on relative performance, reproducibility, and availability of code.
We considered ResViT \cite{dalmaz2022resvit}, I2I-Mamba \cite{atli2024i2imamba}, and dEa-GAN \cite{yu2019ea} as competing GANs, the adversarial DM \textit{SynDiff} \cite{syndiff}, \textit{\textit{SelfRDB}} \cite{arslan2024self} \red{and the DB of Choo et al.~\cite{Cho2024SliceConsistent_MICCAI2024}, and MADM \cite{chen2025multiview}} as competing DMs, and ALDM \cite{kim2024adaptive} as competing 3D LDM. 
All models were trained with the recommended hyperparameters. 
For the training, we adopted the same slice sampling scheme (\red{axial-only}) as \textit{YODA} for all 2D methods (Sec.~\ref{sec:implementation}). 
\textit{SynDiff} was adapted for paired translation \cite{syndiff}. For \red{the DBs} we use the T2w or PD image as the DB prior and added the T1w as an additional conditioning.
\red{We used the 3D Ea-GAN as prior for MADM.} 
\textit{ResViT} was implemented as a single-task model. We halved the size of the VQ-VAE of ALDM \red{and the U-Net of Choo et al.~\cite{Cho2024SliceConsistent_MICCAI2024} for the MRI${\rightarrow}$CT task} to scale the model to our largest hardware (Nvidia A100-40GB).

\subsection{Metrics and statistical analyses} {\setlength{\tabcolsep}{0.35pt} \renewcommand{\arraystretch}{0.15} \newcommand{\imgwidth}{1.8cm}
\begin{large} \begin{figure*}[!b] \centering
\begin{tabular}{ccccccccccccccc}
T1w & T2w  & FLAIR & \emph{diff.} & \(N_\text{Ex}{=}4\)  & \(N_\text{Ex}{=}10\)  & \emph{regr.} &  \\
\includegraphics[width=\imgwidth]{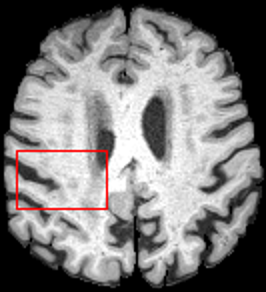} & \includegraphics[width=\imgwidth]{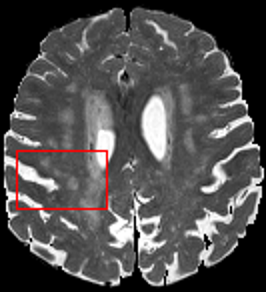} & \includegraphics[width=\imgwidth]{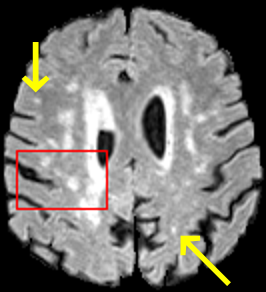} & \includegraphics[width=\imgwidth]{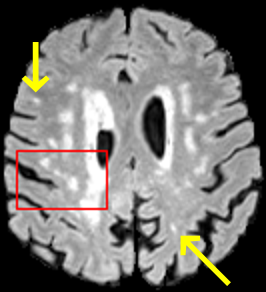} & \includegraphics[width=\imgwidth]{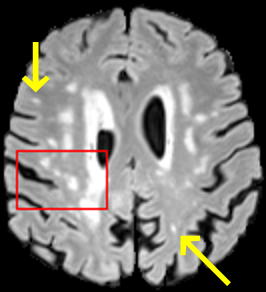} & \includegraphics[width=\imgwidth]{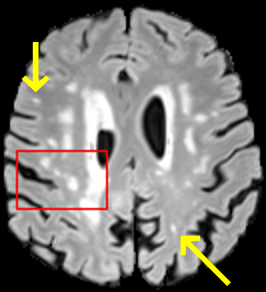} & \includegraphics[width=\imgwidth]{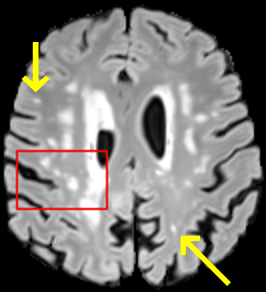} &  \\
\includegraphics[width=\imgwidth]{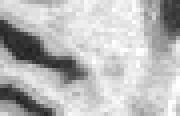} & \includegraphics[width=\imgwidth]{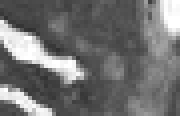} & \includegraphics[width=\imgwidth]{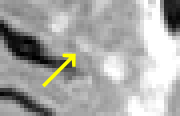} & \includegraphics[width=\imgwidth]{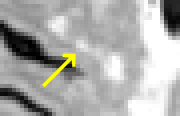} & \includegraphics[width=\imgwidth]{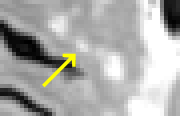} & \includegraphics[width=\imgwidth]{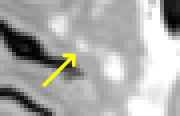} & \includegraphics[width=\imgwidth]{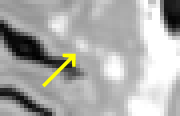}  \\ [1mm]

Choo et al. & MADM & \textit{SelfRDB} & \textit{SynDiff} & ALDM & I2I-Mamba & ResViT & Ea-GAN \\

\includegraphics[width=\imgwidth]{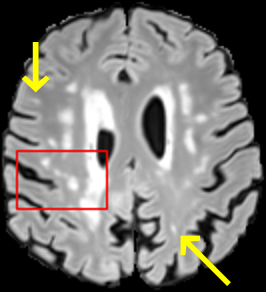} & \includegraphics[width=\imgwidth]{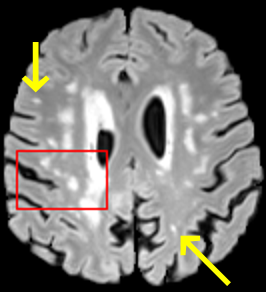} & \includegraphics[width=\imgwidth]{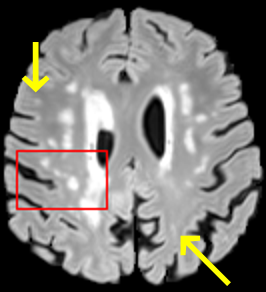} & \includegraphics[width=\imgwidth]{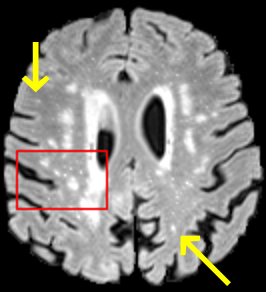} & \includegraphics[width=\imgwidth]{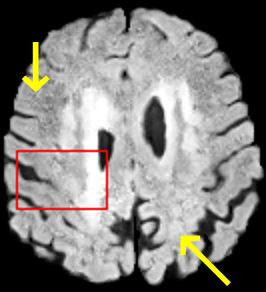} & \includegraphics[width=\imgwidth]{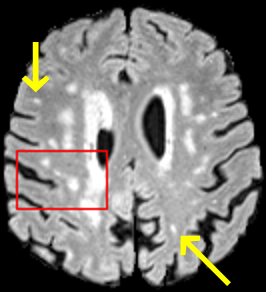} & \includegraphics[width=\imgwidth]{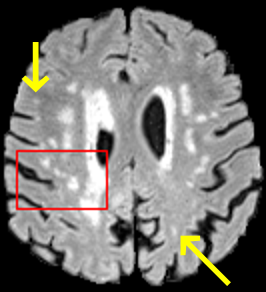} & \includegraphics[width=\imgwidth]{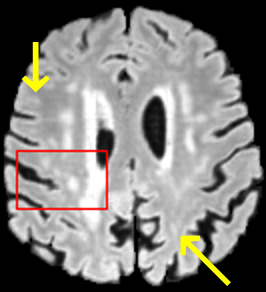} \\
\includegraphics[width=\imgwidth]{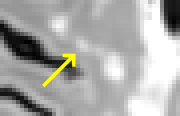} & \includegraphics[width=\imgwidth]{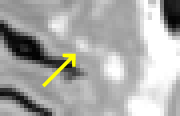} & \includegraphics[width=\imgwidth]{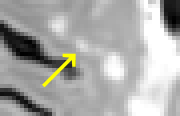} & \includegraphics[width=\imgwidth]{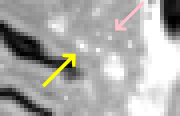} & \includegraphics[width=\imgwidth]{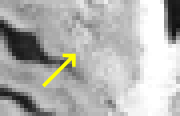} & \includegraphics[width=\imgwidth]{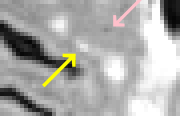} & \includegraphics[width=\imgwidth]{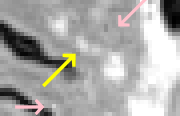} & \includegraphics[width=\imgwidth]{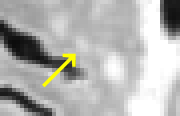} \\ [2mm]

T1w & T2w  & FLAIR & \emph{diff.} & \(N_\text{Ex}{=}4\)  & \(N_\text{Ex}{=}10\)  & \emph{regr.} &  \\

\includegraphics[width=\imgwidth]{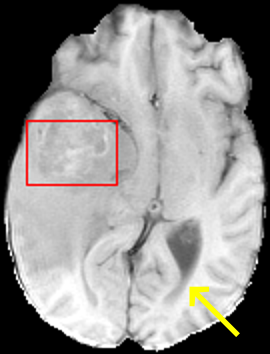} & \includegraphics[width=\imgwidth]{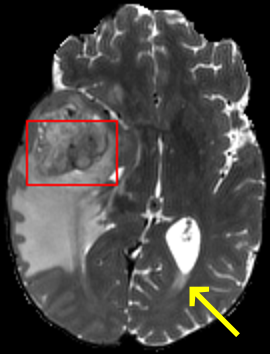} & \includegraphics[width=\imgwidth]{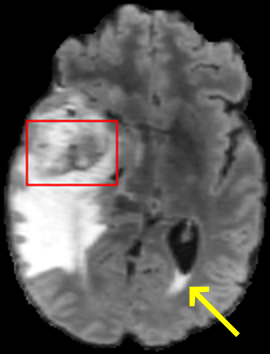} & \includegraphics[width=\imgwidth]{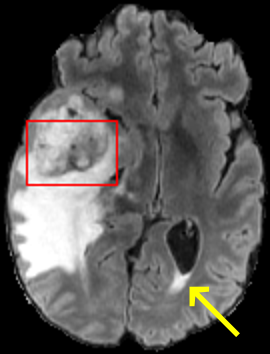} & \includegraphics[width=\imgwidth]{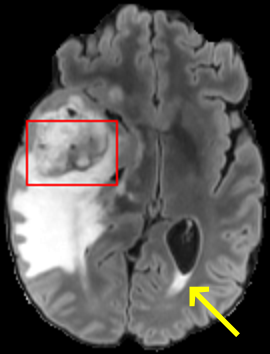} & \includegraphics[width=\imgwidth]{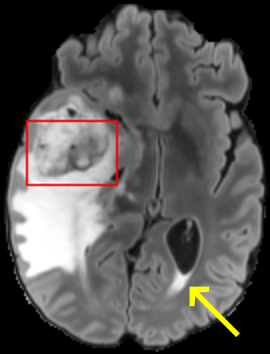} & \includegraphics[width=\imgwidth]{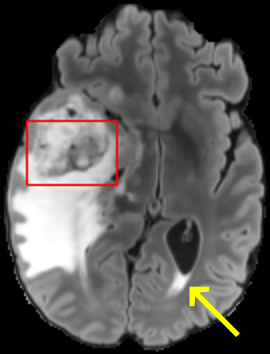} & \\
\includegraphics[width=\imgwidth]{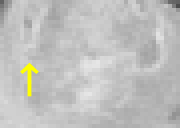} & \includegraphics[width=\imgwidth]{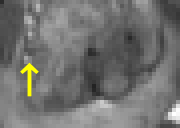} & \includegraphics[width=\imgwidth]{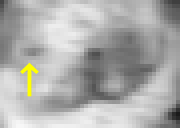} & \includegraphics[width=\imgwidth]{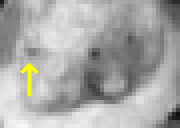} & \includegraphics[width=\imgwidth]{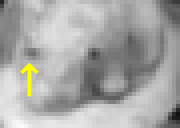} & \includegraphics[width=\imgwidth]{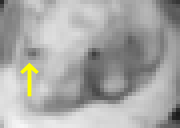} & \includegraphics[width=\imgwidth]{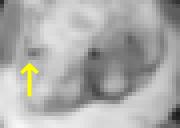}   \\  [1mm]

Choo et al. & MADM & \textit{SelfRDB} & \textit{SynDiff} & ALDM & I2I-Mamba & ResViT & Ea-GAN \\

\includegraphics[width=\imgwidth]{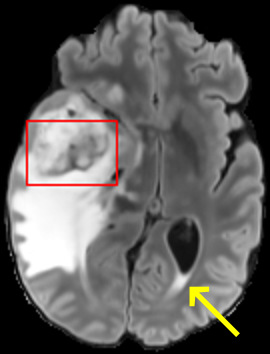} & \includegraphics[width=\imgwidth]{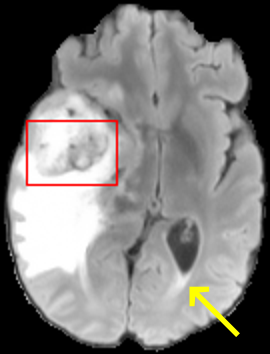} & \includegraphics[width=\imgwidth]{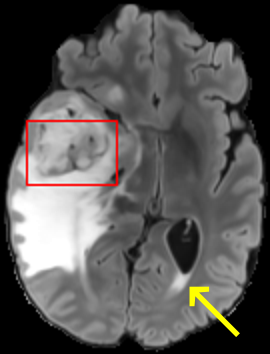} & \includegraphics[width=\imgwidth]{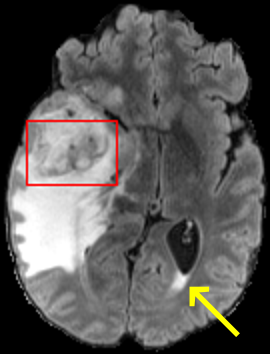} & \includegraphics[width=\imgwidth]{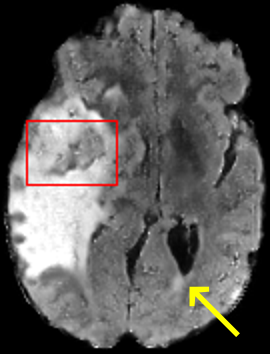} & \includegraphics[width=\imgwidth]{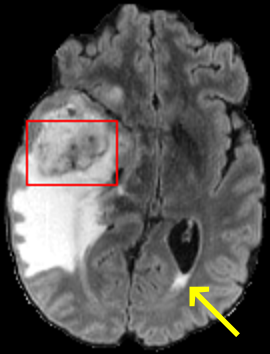} & \includegraphics[width=\imgwidth]{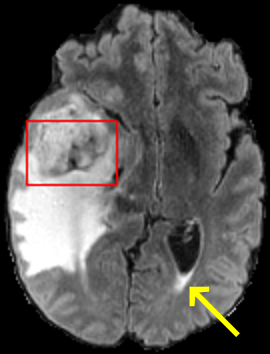} & \includegraphics[width=\imgwidth]{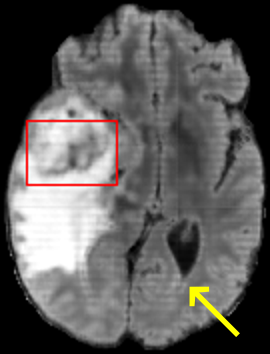} \\
\includegraphics[width=\imgwidth]{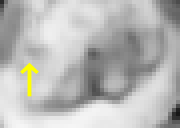} & \includegraphics[width=\imgwidth]{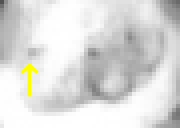} & \includegraphics[width=\imgwidth]{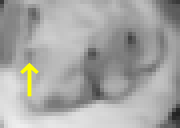} & \includegraphics[width=\imgwidth]{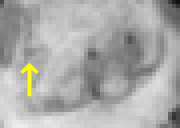} & \includegraphics[width=\imgwidth]{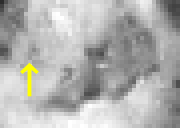} & \includegraphics[width=\imgwidth]{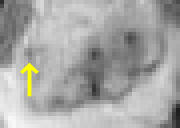} & \includegraphics[width=\imgwidth]{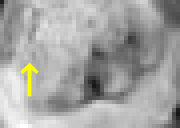} & \includegraphics[width=\imgwidth]{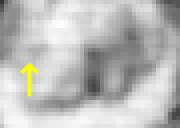} \\

\end{tabular} 
    \caption{Diffusion, ExpA, and \emph{regression} sampling of \textit{YODA} is demonstrated against competing methods for T1w,T2w $\rightarrow$ FLAIR translation on random images of the RS (upper) and BraTS (lower). Note that all sampling methods of \textit{YODA} allow for a more faithful translation of lesions while avoiding artifacts such as Salt-and-Pepper noise and other unrealistic textures. See also the supplementary video for additional translation results.
    } \label{fig:brain_examples} 
\end{figure*} \end{large} }
\subsubsection{Image Metrics} To obtain the 3D SSIM and the PSNR, we applied the brain or pelvic tissue masks and cropped the images to the bounding box.
The perceptual quality was evaluated with the FID ~\cite{fid} using a \textit{ResNet-50} trained on \textit{RadImageNet} \cite{mei2022radimagenet} applied to axial slices. 
\\ 
To estimate the magnitude of noise (\(\hat{\sigma}\) in ~\eqref{eq:noiseMSE}) in synthetic images \(\hat X\), we \red{apply a 3D Gaussian blur (kernel size 5, \(\sigma_\text{blur} = 1.1\)) to the WM voxels and exclude voxel adjacent to tissue boundaries to obtain a smooth intensity map $\bar{X}_\text{WM} \approx X'$. 
We then estimate the noise magnitude \(\bar \sigma _\text{WM}\) as the RMS of the residuals:
\begin{align} \label{eq:wm_noise_eps}
    \bar \sigma_\text{WM} = \text{RMS} (X - \bar X) .
\end{align}}

\subsubsection{Downstream Metrics} For WMH segmentation, we utilized the FLAIR-only version of \textit{SHIVA-WMH} \cite{shiva_wmh} after bias-field correction \cite{n4}. For glioma segmentation, we employed the winner of the 2020 BraTS challenge \cite{isensee2021nnu_general, isensee2021nnu}. In both tasks, we compared the results to the respective reference annotations. For full-brain segmentation from T2w images, we compared \textit{SynthSeg} (v2.0) segmentations \cite{billot2023synthseg} with \textit{FastSurfer} segmentation masks (obtained from T1w images).
For bone and muscle segmentation from CT segmentation, we applied \textit{TotalSegmentator} \cite{wasserthal2023totalsegmentator} relating segmentation from synthetic and acquired images.  
The performance was measured with the Dice score, the Hausdorff distance (HD, 95\textsuperscript{th} percentile), and the absolute log\textsubscript{2} volume ratio (ALVR) \cite{kuijf2019standardized}. 
\red{All downstream tools where used as-is without re-training.}
\\
As a tool-independent measure of WMH translation quality, we define the contrast-to-noise ratio (CNR) for each WMH (i.e.\ each labeled connected component) as
\begin{align} \label{eq:cnr}
    \text{CNR} = \frac{ \mu^{\text{WMH}} - \mu^\text{WM} } { \sigma^\text{WM}},
\end{align}
where \(\mu^{\text{WMH}}\) denotes mean WMH intensity, while \(\mu^\text{WM}\) and \(\sigma^\text{WM}\) denote mean and std.\ of isointense WM within a 3 mm distance from each WMH. We then take voxel-wise averages. 

We report all metrics as mean $\pm$ std. The statistical analyses used non-parametric Wilcoxon signed-rank tests considering \(p < 0.05\) significant without multiple-comparison correction.

\section{Experimental results}
We tested \textit{YODA} for MRI contrast translation in the RS (Tab.~\ref{tab:rs_metrics}, including assessing generalization to the external and distinctive MBB dataset) and BraTS (Tab.~\ref{tab:brats_ixi_metrics}, left) for T1w, T2w $\rightarrow$ FLAIR, on IXI for T1w, PD $\rightarrow$ T2w (Tab. ~\ref{tab:brats_ixi_metrics}, right), and on the pelvic Gold Atlas dataset for MRI $\rightarrow$ CT translation (Tab.~\ref{tab:ct_metrics}). Example images for brain MRI translation are provided in Fig.~\ref{fig:brain_examples} and for pelvis translation in Fig.~\ref{fig:ct_examples}.
First, we analyzed the impact of noise in \textit{YODA}'s sampling methods on the assessed performance and then benchmarked \textit{YODA} performance against competing methods and \red{validated the stability against degraded inputs.}
Finally, we demonstrated the benefits of \textit{YODA} components and design choices in ablation studies.

\subsection{\textit{YODA} sampling methods} \label{sec:perc_dist_trade} 
To test the impact of the iterative refinement on \textit{YODA}'s translation quality, we compared \emph{regression} and \emph{diffusion} sampling on the RS data and present additional generation examples in Fig.~\ref{fig:nex_results}. We observe that \emph{diffusion} sampling visually resembles the appearance of the acquired images. 
\emph{Regression} sampling preserves key anatomical features -- the GM/WM boundary, WMHs (Fig.~\ref{fig:brain_examples}), the outline of the pallidum (Fig.~\ref{fig:nex_results}) -- but omits many high-frequency features. 
To investigate whether iterative refinement during \emph{diffusion} sampling adds relevant and systematic medical information or only imitates acquisition noise, we performed ExpA sampling, i.e.\ averaging the output of several (\(N_\text{Ex}{=}4\) or \(N_\text{Ex}{=}10\)) \emph{diffusion} trajectories.
We observed a gradual loss of high-frequency details when increasing the \(N_\text{Ex}\) (see also the supplementary video), indicating that the effect of the iterative refinement is non-systematic. 
For \(N_\text{Ex}{=}10\), the \emph{images are visually almost indistinguishable from the initial regression solution} \red{(see the supplementary video, and Fig.\ref{fig:brain_examples} and ~\ref{fig:nex_results}). 
We directly compared the synthesis results of ExpA (\(N_\text{Ex}{=}10\)) and \emph{regression} sampling quantitatively and found the differences to be minimal (SSIM: 99.73\%, PSNR: 45.30~dB), i.e.\ diffusion sampling approaches the initial \emph{regression} solution for a high \(N_\text{Ex}\).}
The quantitative analysis of the image quality (Tab.~\ref{tab:rs_metrics}) showed that \emph{diffusion} sampling impairs the assessed SSIM and PSNR in comparison to \emph{regression} sampling for both the in- and external test sets, which we attribute to noise generation (Sec.~\ref{sec:noise_theory}). In turn, ExpA averages improved both metrics and, for \(N_\text{Ex}{=}10\), performed mostly on par with the \emph{regression} solution in both test sets in terms of SSIM, while the PSNR in the RS was slightly increased (Tab.~\ref{tab:rs_metrics}). However, we observed that ExpA sampling \textit{YODA} improves the replication of systematic 3D low-frequency image intensity drifts (bias fields) due to the 3D synchronization in 2.5D \emph{diffusion} sampling. Yet, this apparent advantage did not generalize to the external MBB dataset, as bias fields are MR protocol-specific. 
\subsubsection{The perception-distortion tradeoff}
\red{To further analyze the effect of the backward diffusion process, we analyze how varying the number of iterations $T$ (implicitly $N_\text{FE}$) and noise reduction ($N_\text{Ex}$) affect the assessed performance. Following Blau and Michaeli~\cite{blau2018perception_distortion_tradeoff}, we frame this analysis in the context of the perception-distortion tradeoff, which states that perceptive quality and distortion are opposing objectives.
Here, perceptive quality represents whether the image "looks" similar to the reference image ("image realism") measured by FID. Distortion, on the other hand, quantifies  alignment with the "true" target (PSNR, SSIM) and penalizes not only semantic errors such as hallucinated details but also noise imitation. Increasing $T$ from 1 to 1000 improves preceptive quality at the cost of distortion, increasing $N_\text{Ex}$ reverses this along an almost identical path on the perception-distortion plot (Fig.~\ref{fig:noise_bias_scores}a). Inherently, no method can achieve highest performance in both perceptive quality and distortion (lower left corner in Fig.~\ref{fig:noise_bias_scores}a, see also Tab.~\ref{tab:rs_metrics}) \cite{blau2018perception_distortion_tradeoff}. }

\def\app#1#2{%
  \mathrel{%
    \setbox0=\hbox{$#1\sim$}%
    \setbox2=\hbox{%
      \rlap{\hbox{$#1\propto$}}%
      \lower1.1\ht0\box0%
    }%
    \raise0.25\ht2\box2%
  }%
}
\def\approxprop{\mathpalette\app\relax}

\begin{figure}
{\setlength{\tabcolsep}{0.8pt} \renewcommand{\arraystretch}{0.5} \newcommand{\imgwidth}{1.67cm}
\begin{large}
\centering
\begin{tabular}{ccccc}
acquired & \emph{diff.} & \(N_\text{Ex}{=}10\) & \emph{regr.} & noised \vspace{0.05cm} \\
\includegraphics[width=\imgwidth]{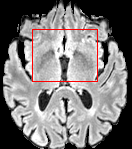} & \includegraphics[width=\imgwidth]{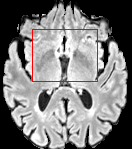} & \includegraphics[width=\imgwidth]{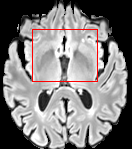} & \includegraphics[width=\imgwidth]{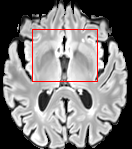} & \includegraphics[width=\imgwidth]{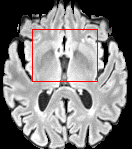} \\
\includegraphics[width=\imgwidth]{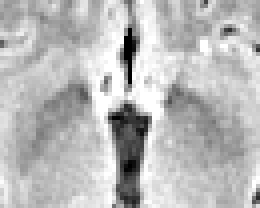} & \includegraphics[width=\imgwidth]{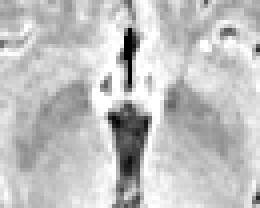} & \includegraphics[width=\imgwidth]{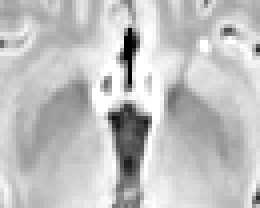} & \includegraphics[width=\imgwidth]{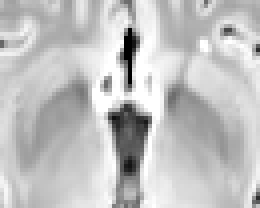} & \includegraphics[width=\imgwidth]{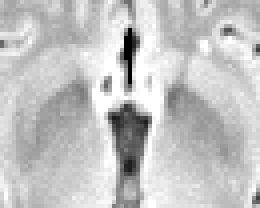} \\
\end{tabular}

    \caption{    
    FLAIR images generated by \textit{YODA} with \emph{diffusion} (\emph{diff.}), ExpA (\(N_\text{Ex}{=}10\)), and \emph{regression} (\emph{regr.}) sampling \red{[...]} are compared to an acquired image (see also the supplementary video).
    Note that ExpA resembles \emph{regression} sampling and achieves crisper edges but looses fine-grained details compared to \emph{diffusion} sampling.
    \red{Additionally, the smooth \emph{regression} image is noised with Rician noise 
    (resulting std.: 1.4\% of $\operatorname{range}(\hat X)$) to restore perceptual realism.}
    } \label{fig:nex_results} 
\end{large} }
    \centering
    \vspace{0.3cm}
    \includegraphics[width=0.99\linewidth]{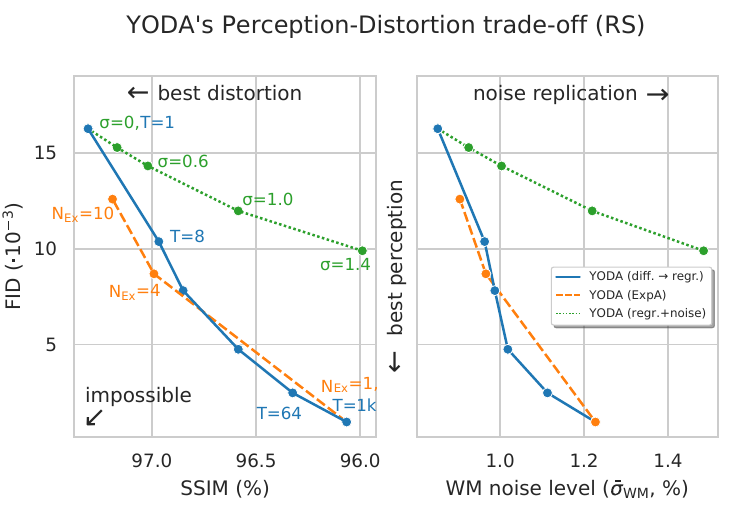}
    \caption{\red{Blau and Michaeli \cite{blau2018perception_distortion_tradeoff} identify that distortion (i.e.\ faithfulness) and perception (i.e.\ realism) are opposing image generation objectives. 
    Here, we show that sliding \textit{YODA}'s sampling parameters \(T\) and \(N_\text{Ex}\) selects the characteristics of generated images on the perception-distortion trade-off curve (see Sec. III-A) in the Rheinland Study (left).
    We transition between \emph{regression} ($T{=}1$, \emph{regr.}) and \emph{diffusion} sampling \textit{YODA} ($T=1k$, \emph{diff.}; both $N_\text{Ex}{=}1$), which gradually improves the perceptual realism (lower FID), but worsens the distortion (lower SSIM). This is reversed by ExpA sampling \textit{YODA} (\(T{=}1k,N_\text{Ex}{=}\{4,10\}\)).
    Conversely, simply adding Rician noise (as resulting std. relative to $\operatorname{range}(\hat X)$) to the smooth \emph{regr.} images (regr.+noise) trades distortion for perceptual realism (Sec.~\ref{sec:noise_theory}).
    Plotting the same configurations as perception over the estimated noise level in the output images (\(\bar\sigma_{\text{WM}} \approxprop \hat \sigma\) in (3)) reveals strong correlation between noise and perception highlighting the role of noise in the perception evaluation (right).
    }}
    \label{fig:noise_bias_scores}
\end{figure}

\red{To better describe the contribution of noise to perceptive quality, we perform two additional analyses: 1.\ We evaluate the effect of simple additive Rician noise at different levels to \emph{regression} prediction of \textit{YODA} ("regr.+noise"). This experiment shows a significant portion of perceptive quality can be recovered by Rician noise, while causing distortion as analyzed theoretically in Section \ref{sec:noise_theory}. 2.\ We plot the FID (perceptive quality) as a function of the "effective noise" (approximated by the standard deviation of WM intensities $\sigma_\text{WM}$, \eqref{eq:wm_noise_eps}) in Fig.~\ref{fig:noise_bias_scores}b and found an inverted correlation: more 
}\red{noise increases perceptive quality.}
\red{
Note that trading distortion} \red{for perception by increasing the number of sampled time steps causes tremendous computational costs, which is only exacerbated when reducing distortion via ExpA sampling (see Tab.~\ref{tab:rs_metrics}).
Yet, to assess whether more realistic, noisy images avoid domain-shifts and to validate if, indeed, more medical information is translated due the iterative refinement \cite{syndiff} but controlling for noise replication in DMs, we consider regular DM and ExpA sampling in the following analyses.}\color{black} 

\subsubsection{Downstream evaluation} \begin{figure}
    \includegraphics[width=0.98\linewidth]{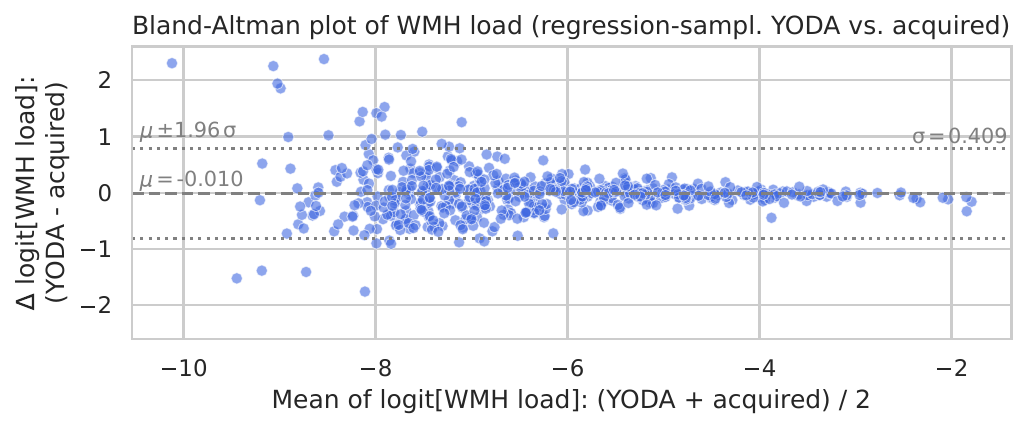}
    \caption{Bland-Altman plot comparing logit-transformed WMH loads obtained from applying the RS WMH pipeline \cite{lohner2022relation} to acquired or \textit{YODA}-generated (\emph{regression} sampling) images.}
    \label{fig:bland_altman}
\end{figure}  \begin{table*}[!b] \centering \addtolength{\tabcolsep}{-0.45em} \begin{scriptsize}
\caption{Performance for brain MRI translation in the BraTS (T1w,T2w $\rightarrow$ FLAIR) and IXI (T1w,PD $\rightarrow$ T2w) datasets. 
\sign \ and \signA \ mark significant differences to \emph{regression} sampling YODA and to segmentation (segm.) on acquired images, respectively. }
\label{tab:brats_ixi_metrics}
\begin{tabular}{l @{\hskip 0.5cm} cc @{\hskip 0.3cm} cc @{\hskip 0.7cm}  cc @{\hskip 0.3cm}  cc}
\toprule
 & \multicolumn{4}{c}{BraTS dataset ($n=200$, downstream glioma segm. \cite{isensee2021nnu})} & \multicolumn{4}{c}{IXI dataset ($n=111$, downstream full-brain segm. \cite{billot2023synthseg})} \\
 & SSIM ($\%, \uparrow$) & PSNR ($\text{dB}, \uparrow$) & Dice ($\%, \uparrow$) & HD ($\text{mm}, \downarrow$) & SSIM ($\%, \uparrow$) & PSNR ($\text{dB}, \uparrow$) & Dice ($\%, \uparrow$) & HD ($\text{mm}, \downarrow$) \\ 
 \midrule
\textit{YODA} (\emph{regr.}) & \textbf{90.87 ± 5.45}\ns & \textbf{27.23 ± 3.19}\ns & \textbf{83.34 ± 14.81}\signA\ns & \textbf{5.13 ± 5.25}\signA\ns & \textbf{98.26 ± 0.73}\ns & \textbf{37.40 ± 2.11}\ns & \textbf{73.54 ± 1.81}\signA\ns & 2.71 ± 0.63\signA\ns \\
\textit{YODA} (\(N_\text{Ex}{=}10\)) & 90.44 ± 5.57\sign & 26.91 ± 3.17\sign & 83.33 ± 14.79\signA\ns & 5.16 ± 5.13\signA\ns & 98.18 ± 0.76\sign & 37.23 ± 2.07\sign & 73.51 ± 1.79\sign\signA & 2.72 ± 0.63\signA\ns \\
\textit{YODA} (\(N_\text{Ex}{=}4\)) & 90.03 ± 5.56\sign & 26.79 ± 3.11\sign & 83.31 ± 14.79\signA\ns & 5.17 ± 5.16\signA\ns & 98.03 ± 0.81\sign & 36.89 ± 2.03\sign & 73.50 ± 1.79\sign\signA & 2.71 ± 0.63\signA\ns \\
\textit{YODA} (\emph{diff.}) & 88.22 ± 5.53\sign & 26.21 ± 2.84\sign & 83.17 ± 14.87\signA\ns & 5.17 ± 5.12\sign\signA & 97.29 ± 1.04\sign & 35.47 ± 1.89\sign & 73.39 ± 1.74\sign\signA & 2.72 ± 0.62\sign\ns \\ 
\midrule
\red{Choo et al.} \cite{Cho2024SliceConsistent_MICCAI2024} &  87.44 ± 5.69\sign &  23.70 ± 3.27\sign  & 82.06 ± 16.41\sign\signA &  5.67 ± 6.33\sign\signA &  97.99 ± 0.83\sign &  35.39 ± 2.27\sign &  73.51 ± 1.88\signA\ns &     2.73 ± 0.62\sign\ns  \\
\red{MADM} \cite{chen2025multiview} &  83.92 ± 5.39\sign &  17.12 ± 2.00\sign & 81.76 ± 14.94\sign\signA &  5.57 ± 5.37\sign\signA  & 94.83 ± 1.31\sign &  29.12 ± 1.25\sign & 73.50 ± 1.68\signA\ns &  \textbf{2.70 ± 0.63}\sign\signA 
\\
\textit{SelfRDB} \cite{arslan2024self} & 88.26 ± 5.45\sign & 25.74 ± 2.66\sign & 82.29 ± 16.31\sign\signA & 5.32 ± 5.16\sign\signA & 97.79 ± 0.86\sign & 35.38 ± 1.85\sign & 73.46 ± 1.72\sign\signA & 2.72 ± 0.61\ns\ns \\
\textit{SynDiff} \cite{syndiff} & 86.56 ± 5.65\sign & 25.19 ± 2.47\sign & 82.13 ± 16.06\sign\signA & 5.31 ± 5.04\sign\signA & 97.38 ± 0.99\sign & 34.69 ± 1.78\sign & 73.00 ± 1.56\sign\signA & 2.82 ± 0.46\sign\signA \\
ALDM \cite{kim2024adaptive} & 77.58 ± 6.08\sign & 21.09 ± 2.55\sign & 80.31 ± 16.33\sign\signA & 5.77 ± 5.32\sign\signA & 86.70 ± 1.96\sign & 25.10 ± 0.97\sign & 62.13 ± 13.13\sign\signA & 4.70 ± 3.43\sign\signA \\
\red{I2I-Mamba} \cite{atli2024i2imamba} & 85.61 ± 5.56\sign &  24.77 ± 2.44\sign & 82.31 ± 16.01\sign\signA &  5.29 ± 5.23\sign\signA &  97.23 ± 1.04\sign &  34.59 ± 1.80\sign & 73.35 ± 1.82\sign\signA &     2.73 ± 0.63\sign\ns
\\
ResViT \cite{dalmaz2022resvit} & 83.35 ± 5.29\sign & 24.14 ± 2.05\sign & 82.43 ± 15.17\sign\signA & 5.30 ± 5.22\sign\signA & 96.64 ± 1.21\sign & 32.82 ± 1.98\sign & 73.51 ± 1.77\signA\ns & 2.72 ± 0.55\sign\ns \\
Ea-GAN \cite{yu2019ea} & 86.24 ± 5.03\sign & 25.00 ± 2.21\sign & 82.14 ± 16.23\sign\signA & 5.41 ± 5.11\sign\signA & 96.69 ± 1.15\sign & 31.91 ± 1.66\sign & 73.50 ± 1.83\signA\ns & 2.71 ± 0.63\ns\ns \\ 
\midrule
Acquired & 100 & - & 84.86 ± 15.14\ns\nsA & 4.90 ± 5.13\ns\nsA & 100 & - & 73.28 ± 1.81\ns\nsA & 2.73 ± 0.63\ns\nsA \\
\bottomrule
\end{tabular}
\end{scriptsize}
\end{table*}
We evaluated the impact of the proposed sampling methods on the performance of the independent, externally trained \textit{SHIVA}-WMH tool as compared to manual reference labels (Tab.~\ref{tab:rs_metrics}). Smoother images (higher \(N_\text{Ex}\) or from \emph{regression} sampling) yielded improved performance over those from \emph{diffusion} sampling.
\red{
In comparison to acquired} \red{images, we noted a} slightly lower Dice score ($p=0.082$), but higher derived volume correlation (ALVR, $p<10^{-3}$) \red{ for \emph{regression}-sampling \textit{YODA}
}(Tab.~\ref{tab:rs_metrics}).
Note that for quality differences as close as reported here, the acquired-to-synthetic comparison is biased in favor of acquired images as the acquisition noise systematically affects both manual and \emph{SHIVA}-WMH segmentations. \\
To assess the suitability of \textit{YODA}-generated images for WMH detection independent of segmentation tools, we also calculated the CNR ~\eqref{eq:cnr} of WMHs. This confirmed that the contrast of WMHs is preserved in the \emph{regression} images, whereas we noted slightly reduced WMH contrast for \emph{diffusion} and ExpA-sampled images (Tab.~\ref{tab:rs_metrics}). \\
As a final assessment, we applied the internal RS WMH segmentation pipeline \cite{lohner2022relation, koch2024rsprotocol} to the smooth \emph{regression} images. As this pipeline was trained with image-label pairs of the RS WMH test, we only compared the derived volumes from inputting acquired vs.\ \emph{regression} images in the larger, unannotated RS test set (\(n{=}599\), one removed due to a processing error). 
Despite the domain shift from noisy to smooth images, the pipeline remained robust and showed excellent agreement (ICC $(3,1)$: 0.982, 95\% CI \([0.98, 0.99]\)) with a near-perfect calibration and only a few outliers (Fig.~\ref{fig:bland_altman}). 

\subsubsection{Confirmation in other datasets}
For BraTS and IXI datasets, we noted the same convergence of visual appearance for \emph{diffusion} to \emph{regression} sampling when increasing the \(N_\text{Ex}\) (Fig. ~\ref{fig:brain_examples}) as observed in the RS. Also, we observed similar tendencies in terms of SSIM and PSNR, and for the performance of the respective downstream tasks (Tab.~\ref{tab:brats_ixi_metrics}).
When assessing segmentation from \textit{YODA}'s synthetic and acquired images, we measured significantly reduced performance for all sampling methods on BraTS, whereas the performance in IXI improved. \\
In the MRI $\rightarrow$ CT translation in the Gold Atlas dataset (Fig.~\ref{fig:ct_examples}), major bones and muscles are translated faithfully. Yet, \emph{diffusion} sampling leads to some inaccuracies in the outline of bones and hallucinations of the textures of inner organs. Conversely, \emph{regression} sampling faithfully translates bones while still blurring the inner organs and generating artifacts. This indicates that rendering CTs is under-defined by the given MRI guidance alone.
In the quantitative analysis (Tab.~\ref{tab:ct_metrics}), we found a slightly increased performance of sampling with \(N_\text{Ex}{=}10\) over \emph{regression} sampling in terms of PSNR and accuracy of the downstream task. As performance gains are marginal and the assessment is based on few ($n=4$) test cases, we still assume the performance of \emph{diffusion} and \emph{regression} to be similar.
\red{We therefore conclude that, beyond slight benefits in fostering 3D coherency, iterative refinement does not have a systematic effect, i.e.\ the expected value of the \emph{diffusion} sampling corresponds to the initial \emph{regression} solution and \emph{diffusion} sampling does not improve image quality beyond perceptual realism.
}
{\setlength{\tabcolsep}{0.35pt} \renewcommand{\arraystretch}{0.15} \newcommand{\imgwidth}{1.45cm}
\begin{figure*}\centering
\begin{tabular}{ccc @{\hskip 0.1cm} ccccccccc}
\multicolumn{1}{c}{\footnotesize T1w} & 
\multicolumn{1}{c}{\footnotesize T2w} & 
\multicolumn{1}{c}{\footnotesize CT} & 
\multicolumn{1}{c}{\footnotesize \emph{diff.}} & 
\multicolumn{1}{c}{\footnotesize \(N_\text{Ex}{=}4\)} & 
\multicolumn{1}{c}{\footnotesize \(N_\text{Ex}{=}10\)} & 
\multicolumn{1}{c}{\footnotesize \emph{regr.}} & 
\multicolumn{1}{c}{\footnotesize Choo et al.} & 
\multicolumn{1}{c}{\footnotesize \textit{SelfRDB}} & 
\multicolumn{1}{c}{\footnotesize \textit{SynDiff}} & 
\multicolumn{1}{c}{\footnotesize I2I-Mamba} & 
\multicolumn{1}{c}{\footnotesize ResViT} \\
\includegraphics[width=\imgwidth]{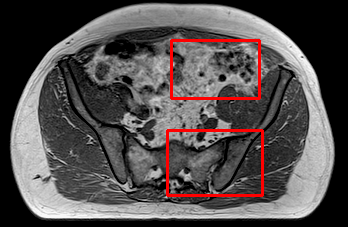} & \includegraphics[width=\imgwidth]{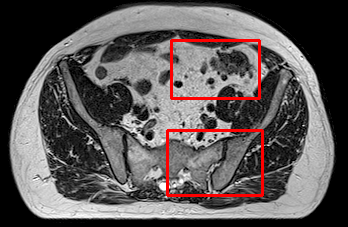} & \includegraphics[width=\imgwidth]{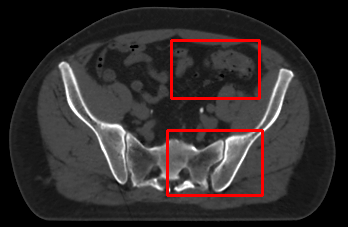} & \includegraphics[width=\imgwidth]{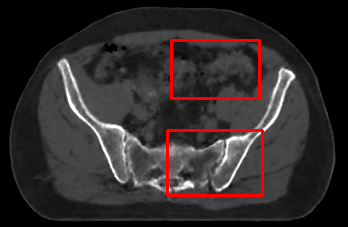} & \includegraphics[width=\imgwidth]{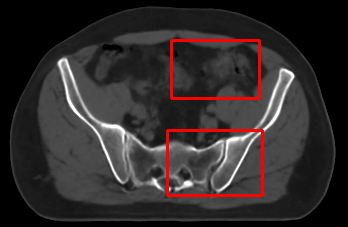} & \includegraphics[width=\imgwidth]{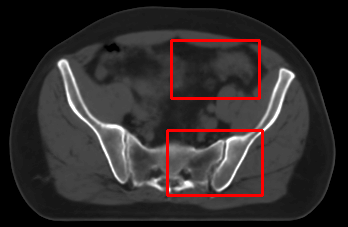} & \includegraphics[width=\imgwidth]{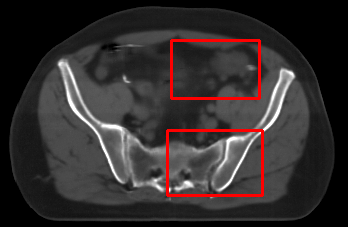} & \includegraphics[width=\imgwidth]{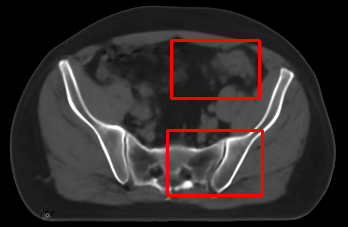} & \includegraphics[width=\imgwidth]{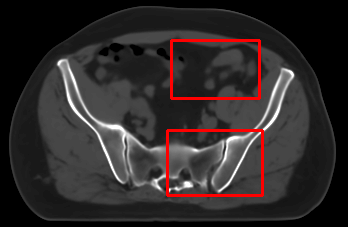} & \includegraphics[width=\imgwidth]{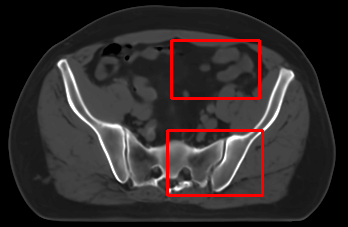} & \includegraphics[width=\imgwidth]{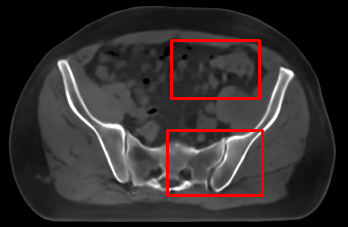} & \includegraphics[width=\imgwidth]{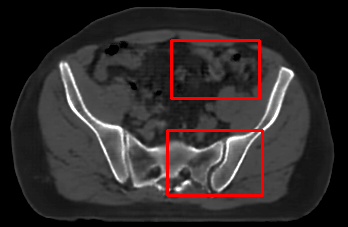} \\
\includegraphics[width=\imgwidth]{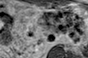} & \includegraphics[width=\imgwidth]{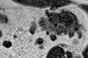} & \includegraphics[width=\imgwidth]{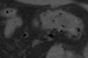} & \includegraphics[width=\imgwidth]{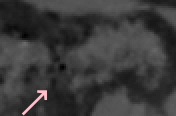} & \includegraphics[width=\imgwidth]{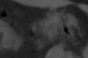} & \includegraphics[width=\imgwidth]{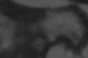} & \includegraphics[width=\imgwidth]{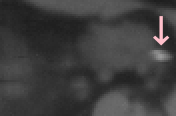} & \includegraphics[width=\imgwidth]{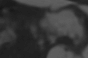} & \includegraphics[width=\imgwidth]{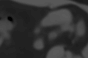} & \includegraphics[width=\imgwidth]{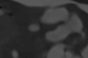} & \includegraphics[width=\imgwidth]{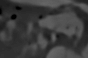} & \includegraphics[width=\imgwidth]{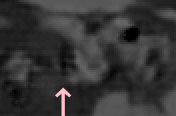} \\
\includegraphics[width=\imgwidth]{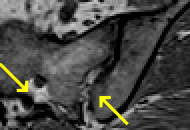} & \includegraphics[width=\imgwidth]{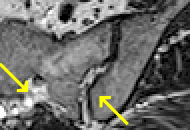} & \includegraphics[width=\imgwidth]{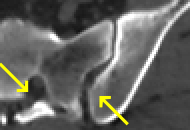} & \includegraphics[width=\imgwidth]{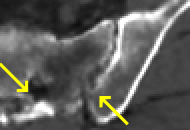} & \includegraphics[width=\imgwidth]{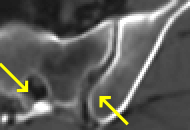} & \includegraphics[width=\imgwidth]{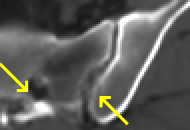} & \includegraphics[width=\imgwidth]{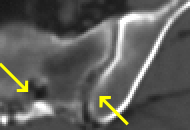} & \includegraphics[width=\imgwidth]{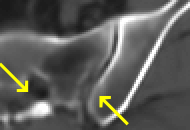} & \includegraphics[width=\imgwidth]{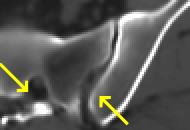} & \includegraphics[width=\imgwidth]{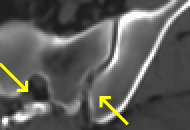} & \includegraphics[width=\imgwidth]{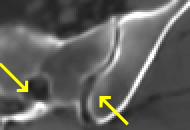} & \includegraphics[width=\imgwidth]{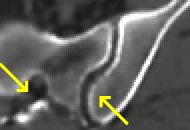} \\
\end{tabular}
    \caption{\textit{YODA} is demonstrated against competing methods for MRI $\rightarrow$ CT translation on the Gold atlas pelvis dataset on a random example. \\
    Diffusion sampling YODA results in hallucinated organ shapes and textures that smoothen out and disappear when increasing the \(N_\text{Ex}\) in Exp sampling or when using regression sampling. Regression sampling creates some artifacts. Note that, for all methods, the translation quality for inner organs is rather poor, whereas bone and muscle translation is reliable. 
    } \label{fig:ct_examples} 
\end{figure*} 
}

\subsection{Performance benchmarking}
\subsubsection{Rhineland Study}
On the RS data (Tab.~\ref{tab:rs_metrics}), \emph{regression} sampling of \textit{YODA} significantly outperformed all competing methods both w.r.t.\ the image quality and \red{-- except the ALVR for MADM --} downstream WMH segmentation, \red{whereas the DB of Choo et al.~\cite{Cho2024SliceConsistent_MICCAI2024} achieved an higher WMH CNR.
\emph{Regression} and ExpA sampling \textit{YODA}} achieved superior \red{SSIM} in the external MBB test set. \\
\red{The FID varied broadly between methods, with \textit{SynDiff} achieving the highest and \emph{diffusion} sampling \textit{YODA} the second} 
\red{highest perceptual quality.}
When analyzing the RS synthesis results (Fig.~\ref{fig:brain_examples}), we noted that most reference methods strive to imitate realistic images, but several artifacts can be observed such as hallucinated WMHs (\textit{SynDiff}) and salt-and-pepper noise (\textit{SynDiff}, \red{I2I-Mamba}, ResViT).
\red{
However, \textit{\textit{SelfRDB}} \cite{arslan2024self}, and, especially, MADM~\cite{chen2025multiview} and the DB of Choo et al.~\cite{Cho2024SliceConsistent_MICCAI2024} generate very smooth images (Fig.~\ref{fig:brain_examples}) with an high FID, i.e.\ low perceptual realism. 
We attributes this to their self-consistent sampling procedures \cite{arslan2024self, Cho2024SliceConsistent_MICCAI2024} and/or accelerated DM sampling with entailed ExpA-like averages during the backward process \cite{Cho2024SliceConsistent_MICCAI2024, chen2025multiview}. 
As such, the increased sampling effort over RMs -- in contrast to \emph{diffusion} sampling \textit{YODA} -- does not result in realistic images, such that these DMs are neither computationally efficient nor do they generate realistic images modeling the distribution of noisy images \(\mathcal{X}\). Thus and in contrast to ExpA sampling \textit{YODA}, these methods deviate from the random process of physical acquisition and signal averages thereof.} 
ALDM is not able to reproduce smaller lesions, which we attribute to missing representations in the compressed latent space. \\
Furthermore, our analysis underlines that pure intensity-based performance metrics are a faulty indicator of WMH translation quality, as e.g., the Ea-GAN achieved competitive SSIM and PSNR, but WMH translation was poor (Tab.~\ref{tab:rs_metrics}). 
\red{Also, artifacts like Salt-and-Pepper noise in competing methods were seemingly not penalized in the FID. In addition, the FID is also blind to slicing artifact (see the supplementary video) by design, as, both, 2D inference and slice-wise feature extraction, are in the axial plane.} 

 \begin{table}
\addtolength{\tabcolsep}{-0.3em}
\begin{scriptsize}
\centering
\caption{Performance for MRI \(\rightarrow\) CT translation in the Gold atlas.}
\label{tab:ct_metrics}
\begin{tabular}{lcc cccc}
\toprule
& \multicolumn{2}{c}{Quality metrics}               & \multicolumn{2}{c}{Bone \& muscle segm.  \cite{wasserthal2023totalsegmentator} }                              \\
& SSIM ($\%, \uparrow$) & PSNR ($\text{dB}, \uparrow$)  & Dice ($\%, \uparrow$)  & HD ($\text{mm}, \downarrow$)  \\ 
\midrule
\textit{YODA} (\emph{regr.}) & 89.47 ± 2.01 & 25.46 ± 1.41 & 94.40 ± 1.23 & 2.01 ± 0.41 \\
\textit{YODA} (\(N_\text{Ex}{=}10\)) & \textbf{89.64 ± 1.96} & 25.32 ± 1.28 & \textbf{94.82 ± 1.19} & 1.92 ± 0.41 \\
\textit{YODA} (\(N_\text{Ex}{=}4\)) & 88.74 ± 1.82 & 25.12 ± 1.28 & 94.37 ± 0.99 & 2.00 ± 0.33 \\
\textit{YODA} (\emph{diff.}) & 87.67 ± 1.76 & 25.00 ± 1.24 & 94.32 ± 0.91 & 2.01 ± 0.31 \\
\midrule
\red{Choo et al.} \cite{Cho2024SliceConsistent_MICCAI2024} & 89.32 ± 1.42 &  \textbf{26.50 ± 1.83} & 94.45 ± 0.84 &  1.87 ± 0.19 \\
\textit{SelfRDB} \cite{arslan2024self} & 87.87 ± 1.29 & 24.91 ± 1.37 & 92.92 ± 1.28 & 2.29 ± 0.19 \\
\textit{SynDiff} \cite{syndiff}& 86.98 ± 1.51 & 24.54 ± 1.08 & 93.76 ± 1.03 & 2.10 ± 0.31 \\
\red{I2I-Mamba} \cite{atli2024i2imamba} & 88.46 ± 1.73 &  25.62 ± 1.97 & 94.43 ± 1.18 & \textbf{1.83 ± 0.29}\\
ResViT \cite{dalmaz2022resvit} & 85.16 ± 3.87 & 24.12 ± 1.39 & 93.57 ± 2.24 & 2.10 ± 0.56 \\
\bottomrule
\end{tabular} \end{scriptsize}
\end{table}
\red{To isolate the effects of the proposed \emph{regression} sampling paradigm from \textit{YODA}'s remaining 2.5D components (multi-slice inputs, view aggregation), we additionally benchmarked a smaller, 2D-only \textit{YODA} version on the RS.
Among all assessed methods, the smaller \textit{YODA} has the second lowest resource requirements in terms VRAM usage and training time (Tab.~\ref{tab:resources}) and fastest inference time (Tab.~\ref{tab:rs_metrics}).
Yet, this version still outperformed all competing methods w.r.t.\ to SSIM, except the DB of Choo et al.~\cite{Cho2024SliceConsistent_MICCAI2024} (Tab.~\ref{tab:rs_metrics}). Note that this DM uses 3D information \cite{Cho2024SliceConsistent_MICCAI2024} and employs a larger backbone model (Tab.~\ref{tab:resources}), which overcompensates the weak noise remaining after accelerated sampling and averaging. 
In turn, the larger 214M \textit{YODA} version is still competitive in terms of training requirements and inference time.
Note that the \emph{regression} inference of \textit{YODA}} is faster than FLAIR acquisition (Tab.~\ref{tab:resources}). 

\begin{table}[!b]
\centering 
\caption{
\red{
Comparison of resource requirements and model size. 
Training effort on NVIDIA V100s (ALDM: A100s) for the RS dataset (1344 volumes). The parameter counts are separated for the Generator/Discriminator (G/D, GANs) or VAE/U-Net (V/U, LDM).  
}}
\label{tab:resources}
\begin{scriptsize}
\begin{tabular}{lccc}
\toprule
\multirow{2}{*}{method} & training effort      & training (inference)     & \multirow{2}{*}{\# parameters} \\
                        & $[\text{GPU hours}]$ & VRAM usage $[\text{GB}]$ & \\
\midrule
\textit{YODA} (large) & 670 & 5.5 (2.3) & 214M \\
\textit{YODA} (small) & 140 & 2.7 (1.8) & 54M \\
Choo et al. & 1220 & 14.6 (11.7) & 254M \\
MADM  & 1365 & 6.0 (4.6) & $3 \times 140$M \\  
\textit{SelfRDB} & 930 & 4.9 (1.4) & 40M (G), 7M (D) \\
\textit{SynDiff} & 550 & 4.2 (1.1) & 40M (G), 18M (D) \\
\textit{ResViT} & 200 
& 3.3 (1.1) & 123M (G), 3M (D) \\
I2I-Mamba & 280 
& 2.6 (1.8) & 109M (G), 3M (D)  \\
\midrule
ALDM & 910 & 40 (22) & 9M (V), 136M (U) \\
Ea-GAN & 45 & 6.5 (4.4) & 17M (G), 11M (D) \\ 
\bottomrule
\end{tabular} \end{scriptsize}
\end{table}

\subsubsection{Confirmation in other datasets}
When benchmarking on the more heterogeneous and lower-quality BraTS dataset (Tab.~\ref{tab:brats_ixi_metrics}, left), \emph{regression} sampling of \textit{YODA} achieved significantly better images than all competing methods in terms of PSNR, SSIM, and downstream glioma segmentation. Similar results were obtained in the IXI dataset (Tab.~\ref{tab:brats_ixi_metrics}, right), where \textit{YODA} significantly outperformed all competing methods in SSIM and PSNR. In the full-brain segmentation conducted from the synthetic images, \textit{YODA}'s images were significantly better than all competing methods in at least either the Dice score or the HD. 
In the MRI \(\rightarrow\) CT task on the small pelvic dataset, \emph{regression} and ExpA sampling achieved the highest SSIM, 
\red{whereas the DB of Choo et al.~\cite{Cho2024SliceConsistent_MICCAI2024} had the highest PSNR.
Both, \emph{regression} and ExpA sampling \textit{YODA} also performed competitively in the downstream segmentation} \red{task.}
Note that, for all methods, the translation of inner organs was rather poor and included either texture hallucination \red{(ResViT, I2I-Mamba, \emph{diffusion}-sampling \textit{YODA}) or an over-}\red{smooth image appearance (Choo et al.~\cite{Cho2024SliceConsistent_MICCAI2024}, \textit{SelfRDB}, \textit{SynDiff}, ExpA and \emph{regression} sampling \textit{YODA}, Fig.~\ref{fig:ct_examples}). 
}

\subsection{Analyzing robustness}
\subsubsection{Worst performance cases} {\setlength{\tabcolsep}{0.35pt} \renewcommand{\arraystretch}{0.15} \newcommand{\imgwidth}{1.990cm}
\begin{large} \begin{figure}\centering
\begin{tabular}{cccc}
T1w & T2w & FLAIR & \textit{YODA} (\emph{regr.}) \\
\includegraphics[width=\imgwidth]{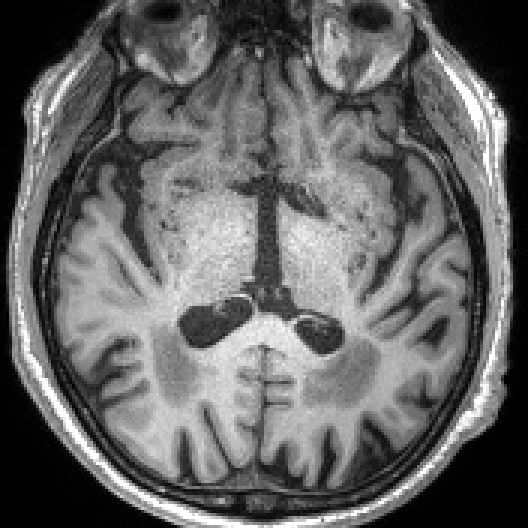} & \includegraphics[width=\imgwidth]{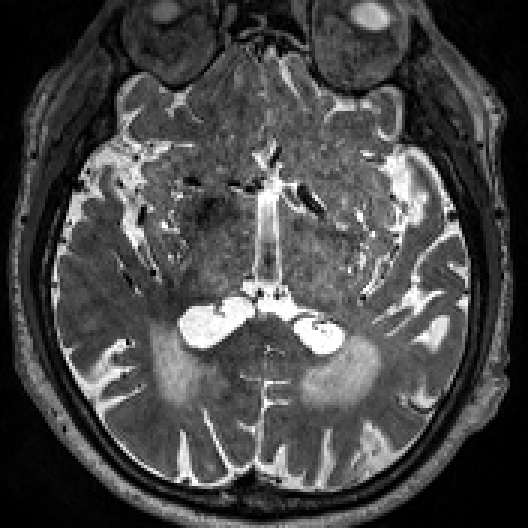} & \includegraphics[width=\imgwidth]{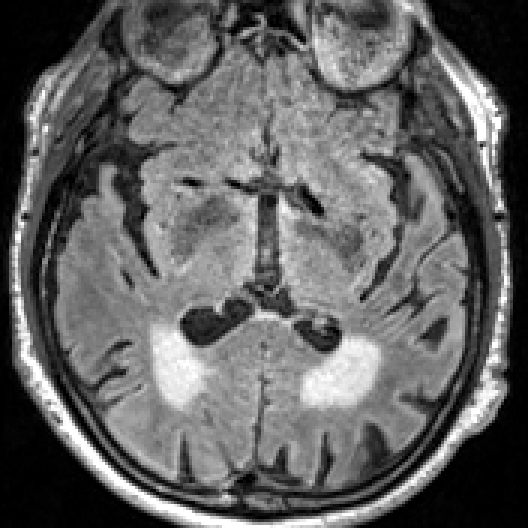} & \includegraphics[width=\imgwidth]{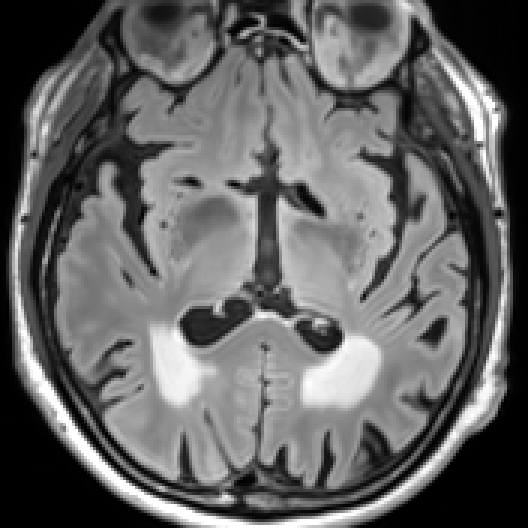} \\
\includegraphics[width=\imgwidth]{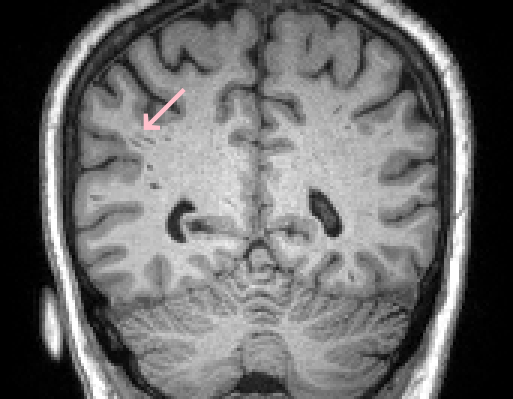} & \includegraphics[width=\imgwidth]{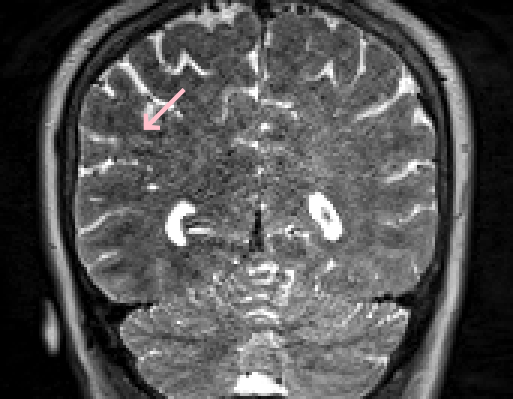} & \includegraphics[width=\imgwidth]{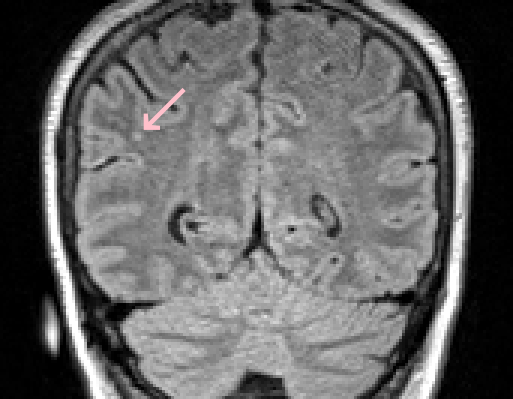} & \includegraphics[width=\imgwidth]{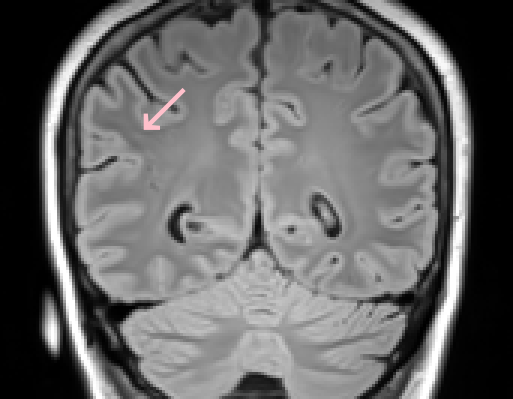} \\
\includegraphics[width=\imgwidth]{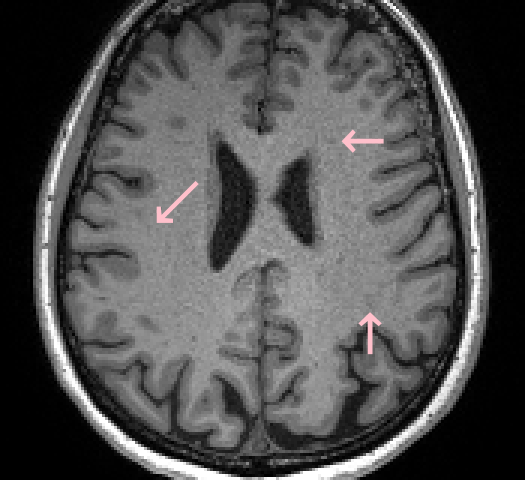} & \includegraphics[width=\imgwidth]{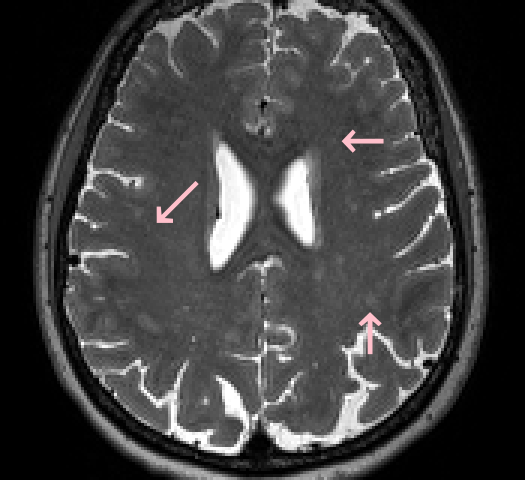} & \includegraphics[width=\imgwidth]{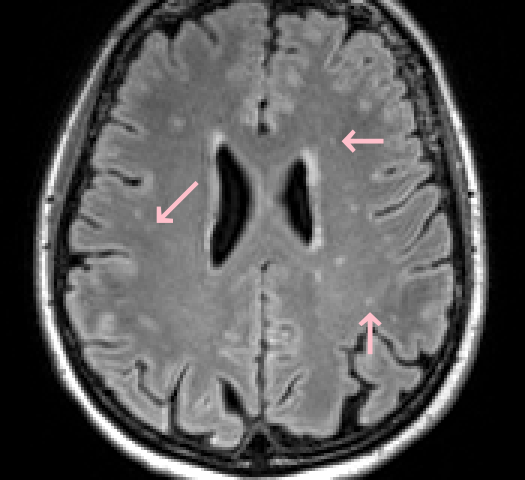} & \includegraphics[width=\imgwidth]{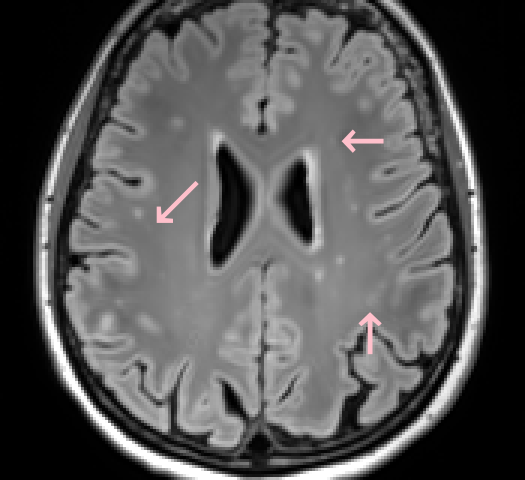} \\ [.5mm]
\includegraphics[width=\imgwidth]{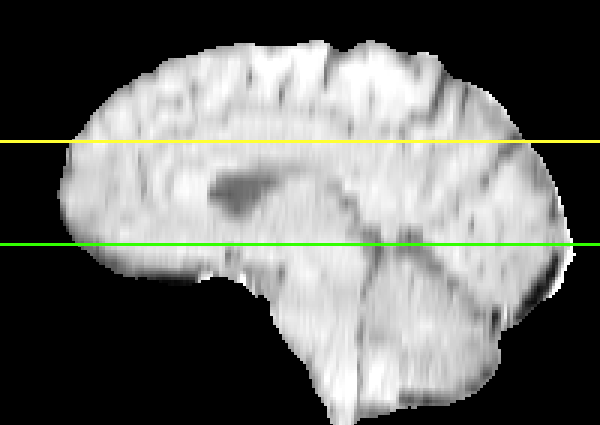} & \includegraphics[width=\imgwidth]{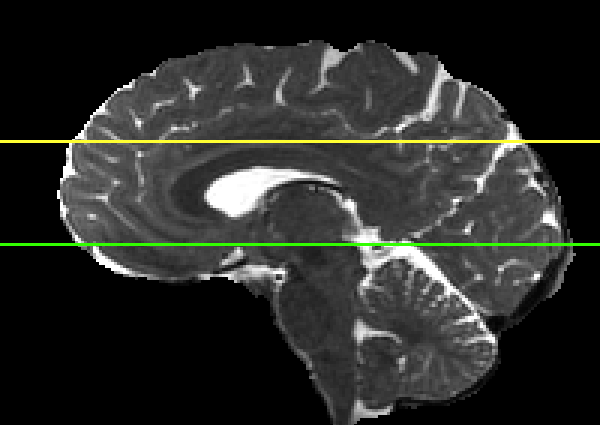} & \includegraphics[width=\imgwidth]{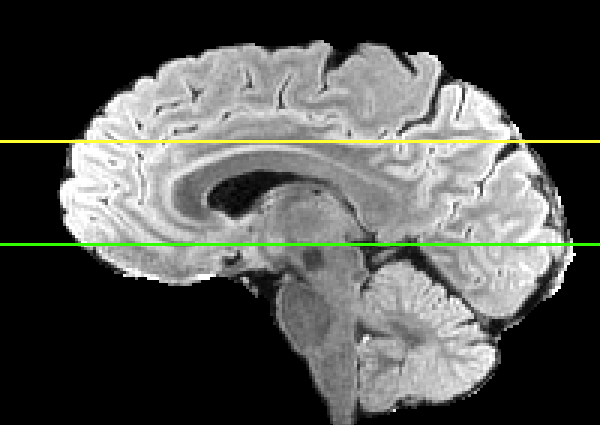} & \includegraphics[width=\imgwidth]{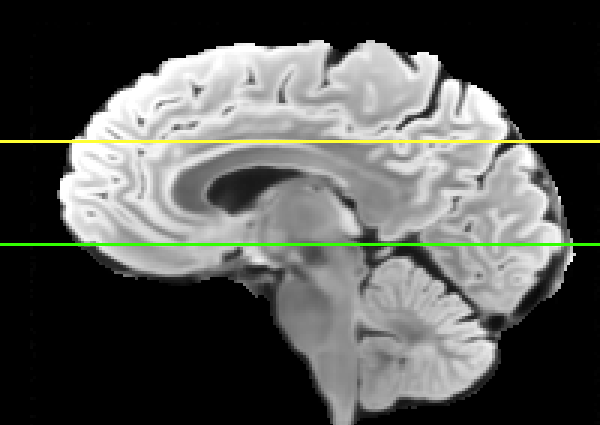} \\
\includegraphics[width=\imgwidth]{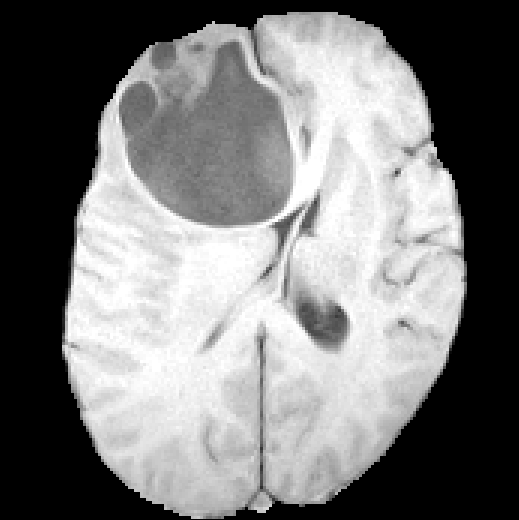} & \includegraphics[width=\imgwidth]{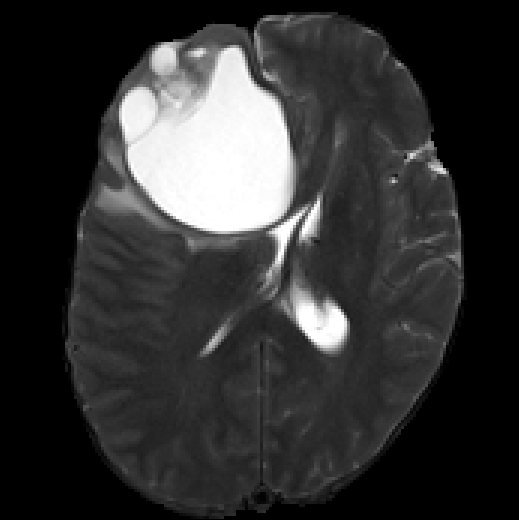} & \includegraphics[width=\imgwidth]{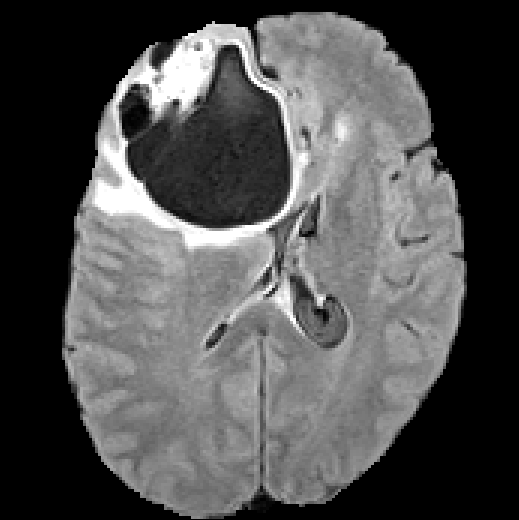} & \includegraphics[width=\imgwidth]{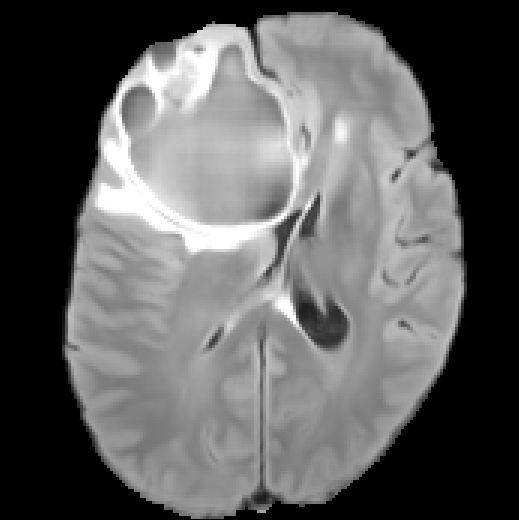} \\
\end{tabular}
\caption{\red{Representative worst synthesis results for \emph{regression} sampling \textit{YODA}.
Typical error patterns in the RS were reduced source and target image quality (row 1 and 2) and difficulties in translating small WMHs (row 2 and 3), whereas bad spatial alignment of source and target images impaired synthesis quality in BraTS (row 4). Row 5 shows the severest semantic error in which the signal of Fluid-filled cavity was not suppressed in the synthetic FLAIR images.
}} \label{fig:worst_case}
\end{figure}  \end{large} }
\red{As we could not find clear patterns of bad performance in random inspection and did not observe hallucination, we visually inspected the 10\% of worst translation results w.r.t.\ SSIM and downstream Dice score (at most 10 images) in the RS, BraTS, and IXI (total: $n=55$ images ) and show representative cases in Fig.~\ref{fig:worst_case}.
We observed that the low performance was mainly caused by deteriorated target (first row, Fig.~\ref{fig:worst_case}) or source images (second row, Fig.~\ref{fig:worst_case}) or registration errors (especially Brats, forth row, Fig.~\ref{fig:worst_case}) or combinations of these factors.
Thus, this strategy led to an effective quality control to filter for low-quality acquisitions, i.e.\ detecting images with large $\sigma$ in ~\eqref{eq:noiseMSE}).
However, we observed some missing small WMHs (third row, Fig.~\ref{fig:worst_case}) in the synthetic FLAIR images. Due to the weak input signal, translating small WMHs in inherently difficult and, thus, omitting these WMHs can be seen as conservative. In contrast, a more speculative translation bears the risk of false-positive, hallucinated WMHs (compare e.g., \textit{SynDiff}, Fig.~4). \\    
In Brats, we observed one case with FLAIR hyper- rather than hypo-intense tumor cavity (Fig.~\ref{fig:worst_case}, fifth row). 
While this constitutes a semantic error suggesting soft-tissue rather than fluids, we note that the differentiation from T1w and T2w images alone is ambiguous.}

\subsubsection{MIT from degraded inputs} \begin{figure} \vspace{-2mm}    \centering
    \includegraphics[width=0.91\linewidth]{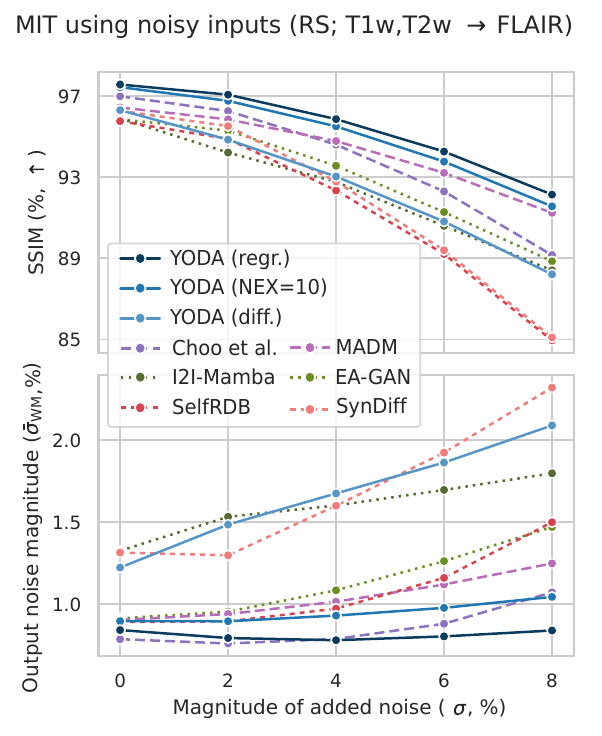}
    \caption{\red{
    The stability to noisy inputs is assessed for \textit{YODA} and competing methods by adding varying magnitudes of Rician noise in a subset ($n{=}50$) of the RS test set.
    The resulting noise std. ($\sigma$ as \% of $\operatorname{range}(C)$) added to the input images is plotted against the SSIM, PSNR and estimated WM noise levels in the synthetic images. 
    }}
    \label{fig:noise_input}
\end{figure}
\red{
As an additional validation, we assessed the stability against noisy inputs by adding Rician noise of various magnitudes to the source images (Fig.~\ref{fig:noise_input}).
Note that noise levels of the source and target images are correlated due to, among others, motion affecting both co-acquired images similarly \cite{pollak2023quantifying} (inherent noise correlation). 
To avoid the confounding effect of deteriorated reference quality when evaluating real low-quality input images, we resorted to simulated noise.
Irrespective of the sampling method, the image quality of \textit{YODA} remained robust against even unrealistically high levels of noise and, thus, out-of-distribution inputs.
In fact, ExpA and \emph{regression} sampling \textit{YODA} achieved a higher SSIM than all competing methods at all noise levels. 
Moreover, these generated images consistently featured a low-noise appearance (e.g., in the WM $\bar \sigma_\text{WM}$, Fig.~\ref{fig:noise_input} bottom row). In contrast, for images generated by \textit{diffusion} sampling \textit{YODA} (but also GAN-based I2I-Mamba, Ea-GAN, and the adversarial DMs \textit{SynDiff} and \textit{SelfRDB}), we observed that higher input noise levels led to increased noise replication.
We attribute this correlation of input and output noise magnitudes to replicating patterns from the training data (i.e.\ the inherent noise correlation).
MADM and the DB of Choo et al.~\cite{Cho2024SliceConsistent_MICCAI2024} did not increase noise replication for low and moderate noise $\sigma \leq 4$ added to the inputs, i.e.\ these DMs do not imitate the correlated noise levels from the training set.
Contrary to the other GANs, the synthesis quality of I2I-Mamba remained robust against input noise achieving the best PSNR for \(\sigma \geq 6 \), while noise replication increased steadily.}

 {\setlength{\tabcolsep}{0.8pt} \renewcommand{\arraystretch}{0.5} \newcommand{\imgwidth}{2cm}
\begin{large}
\begin{figure}[b!]
\centering
\begin{tabular}{ccc}
\includegraphics[width=\imgwidth]{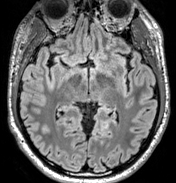} & \includegraphics[width=\imgwidth]{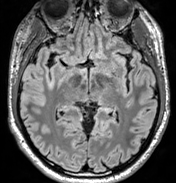} & \includegraphics[width=\imgwidth]{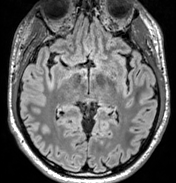} \\[0.7mm]
\includegraphics[width=\imgwidth]{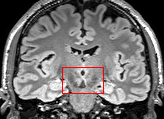} & \includegraphics[width=\imgwidth]{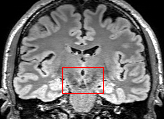} & \includegraphics[width=\imgwidth]{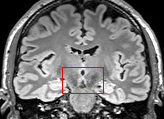} \\ 
\includegraphics[width=\imgwidth]{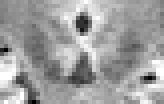} & \includegraphics[width=\imgwidth]{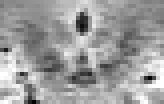} & \includegraphics[width=\imgwidth]{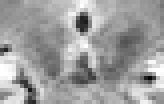}  \\[0.7mm]
\includegraphics[width=\imgwidth]{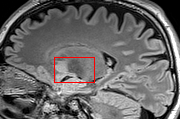} & \includegraphics[width=\imgwidth]{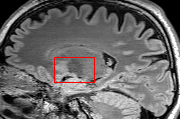} & \includegraphics[width=\imgwidth]{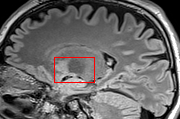} \\
\includegraphics[width=\imgwidth]{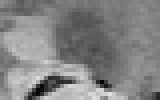} & \includegraphics[width=\imgwidth]{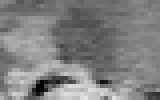} & \includegraphics[width=\imgwidth]{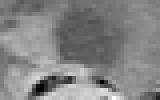} \\
\end{tabular}

    \caption{    
    \red{Left to right: Acquired RS image, and \emph{diffusion}-sampled images from 2D \red{(axial only)} or 2.5D \textit{YODA} versions, respectively.}
    } \label{fig:abl_25D} 
\end{figure} \end{large} } \begin{table} \centering
\addtolength{\tabcolsep}{-0.3em}
\begin{scriptsize}
\centering
\caption{Ablation of \red{number of input slices (\# inp. sl.) } and \red{inference methods} on the RS dataset.
\quad  \sign \ significant difference to a \red{single-slice, axial-only} model.}
\label{tab:abl_25D}
\begin{tabular}{lc@{\hskip 0cm}c @{\hskip 0.3cm} cc}
\toprule
sampling method             & \# inp. sl. & 3D component & SSIM ($\%, \uparrow$) & PSNR (dB, $\uparrow$)\\
\midrule
\multirow{5}{*}{\emph{diffusion}} & \red{9}   & \red{orthogonal denoising}   & \textbf{95.92 ± 1.34}\sign &  33.09 ± 1.36\sign \\
                       & 5   & orthogonal denoising   & \textbf{95.92 ± 1.34}\sign & \textbf{33.13 ± 1.33}\sign \\ 
                       & 1   & orthogonal denoising   & 95.85 ± 1.31\sign & 33.09 ± 1.31\sign \\
                       & 5   & axial only   & 95.86 ± 1.30\sign & 32.88 ± 1.24\sign \\
                       & 1   & axial only   & 95.27 ± 1.35\ns & 32.23 ± 1.16\ns \\
                       \midrule
\multirow{4}{*}{\emph{regression}}  & \red{5}  & \red{view aggr.}& \textbf{97.27 ± 1.11}\sign & \textbf{34.76 ± 1.58}\sign \\
                        & \red{9}  & \red{axial only}  & 97.12 ± 1.12\ns & 34.33 ± 1.45\ns  \\
                        & 5  & axial only  & 97.12 ± 1.12\ns & 34.29 ± 1.51\ns \\ 
                        & 1  & axial only  & 97.11 ± 1.12\ns & 34.23 ± 1.54\ns \\
\bottomrule
\end{tabular}
\end{scriptsize}
\end{table} 

\subsection{Ablation of \textit{YODA} components} 
\subsubsection{2.5D components} 
\textcolor{black}{To enable coherent 3D sampling of 2D DMs, we introduced 2.5D diffusion.
Single-slice and single-plane denoising results in slicing artifacts due to diverging sampling trajectories in adjacent slices, which are prevented by multi-slice inputs and orthogonal denoising (Fig.~\ref{fig:abl_25D}). Both 2.5D modifications individually resulted in significantly improved SSIM and PSNR, while combining both led to additional improvements (Tab.~\ref{tab:abl_25D}).} \red{Multi-slice inputs improved \emph{regression} sampling only slightly, whereas the additional view aggregation led to a significantly improved performance.}

\subsubsection{Noise schedule} 
\color{black}
We exchanged \textit{YODA}'s default linear for a cosine noise schedule, which shifts the diffusion process towards higher SNRs.
For both \emph{diffusion} and \emph{regression} sampling, this significantly decreased SSIM and PSNR, whereas the FID improved (Tab.~\ref{tab:abl_sampl}), i.e.\ perceptual quality is traded for more severe distortion.
We attribute impaired \emph{regression} quality to reduced training effort at low SNRs, i.e.\ image translation without meaningful diffusion priors. 
Nevertheless, both schedules lead to similar image quality under ExpA sampling, which implies that the cosine schedule simply shifts the translation towards higher SNRs without impacting the overall translation capabilities.
Preferring the computationally cheaper \emph{regression} solution, we disregarded the cosine schedule. \\
Based on the quality and stability of the predicted images \(\hat X_{t\rightarrow0}\) at early diffusion steps ($t \gg 0$, i.e.\ low SNR) when using the linear schedule, we adopted a truncated sampling scheme skipping the first $3/4$ steps to spare compute \red{(Fig.~\ref{fig:trunc})}. This does not have any measurable impact on image quality in comparison to running the full reverse process (Tab.~\ref{tab:abl_sampl}).

\subsubsection{Training dedicated RMs}
\red{To further explore the effect of focusing model training on single-shot image generation without diffusive priors, we additionally trained a DM with \(T{=}1\), i.e.\ a dedicated RM (last row in Tab.~\ref{tab:abl_sampl}).
This slightly improved regressive image generation over \emph{regression}-sampling \textit{YODA} (linear schedule). Yet, we did not train dedicated RMs to avoid training extra models for \emph{diffusion} and ExpA sampling and for a fair comparison of sampling paradigms by using the same weights.
Note that this experimental design also demonstrates that DMs know the best, noise-free solution, but deliberately worsen image quality during the iterative refinement (Fig.~\ref{fig:overview}).
}

\subsubsection{Prediction target}  \begin{table} \centering
\addtolength{\tabcolsep}{-0.3em}
\begin{scriptsize}
\caption{
Ablation of noise schedules (cosine vs.\ truncated and full linear) on the RS dataset. 
\red{ \; ¹ $\cdot 10^{-3}$, a.u.}
\red{\; ² corresponds to a regular RM.}
\quad \sign \  significant differences to the default setting (truncated linear schedule)
}
\label{tab:abl_sampl}
\begin{tabular}{llccc}
\toprule
sampling & schedule & \multirow{2}{*}{SSIM ($\%, \uparrow$)} & \multirow{2}{*}{PSNR (dB, $\uparrow$)} & \multirow{2}{*}{FID¹ ($\downarrow$)} \\
method & (train and test) &  &  &  \\ 
\midrule
\multirow{3}{*}{\emph{diffusion}} & lin. (truncated) & \textbf{95.92 ± 1.34}\ns & \textbf{33.14 ± 1.32}\ns & 1.06 \\
    & lin. (full) & \textbf{95.92 ± 1.34}\ns & 33.13 ± 1.33\ns & 1.05 \\
    & cos. & 95.83 ± 1.37\ns & 33.02 ± 1.35\ns & \textbf{0.76} \\ \midrule
\multirow{3}{*}{ExpA (\(N_\text{Ex}{=}4\))} & lin. (truncated) & 96.92 ± 1.18\ns & 34.34 ± 1.49\ns & 11.94 \\
    & \red{lin. (full)} &  \textbf{96.94 ± 1.17}\ns &    \textbf{34.37 ± 1.50}\ns & 12.07 \\
     & cos. & 96.92 ± 1.19\ns & 34.29 ± 1.53\ns & \textbf{11.48} \\ 
\midrule
\multirow{3}{*}{\emph{regression}} & lin. & 97.12 ± 1.12\ns & 34.29 ± 1.51\ns & 20.53 \\
     & cos. & 96.76 ± 1.12\sign & 33.85 ± 1.46\sign & \textbf{17.54} \\ 
    & single-step (\(T{=}1\))² & \textbf{97.14 ± 1.10}\sign & \textbf{34.38 ± 1.48}\sign & 20.54 \\
\bottomrule
\end{tabular} \end{scriptsize}
\end{table} \begin{table}[b!] \centering
\addtolength{\tabcolsep}{-0.3em}
\begin{scriptsize}
\caption{
Ablation of velocity (\(v\)), noise (\(\varepsilon\)), and image (\(X_0\)) prediction on the RS dataset. \  \sign \  significant differences to \(v\)-prediction.}
\label{tab:abl_pred_target}
\centering
\begin{tabular}{lccc}
\toprule
sampl. & target &  SSIM ($\%, \uparrow$) & PSNR (dB, $\uparrow$)   \\
\midrule
\multirow{3}{*}{diff.} & \(v\)   & \textbf{95.97 ± 1.13}\ns & \textbf{33.16 ± 1.28}\ns\\
    &\(\varepsilon\)   & 95.67 ± 1.21\sign & 32.79 ± 1.33\sign  \\ 
     & \(X_0\)  & 94.94 ± 1.15\sign & 31.91 ± 1.09\sign \\ \midrule
\multirow{3}{*}{ExpA (\(N_\text{Ex}{=}4\))} & \(v\)       & \textbf{96.97 ± 0.96}\ns & \textbf{34.36 ± 1.44}\ns  \\
    & \(\varepsilon\)   & 96.77 ± 0.99\sign & 34.07 ± 1.44\sign \\
    & \(X_0\)  & 96.23 ± 0.98\sign & 33.14 ± 1.22\sign \\ \midrule
\multirow{2}{*}{regr.} & \(v\)       & \textbf{97.16 ± 0.90}\ns & \textbf{34.31 ± 1.47}\ns \\
     & \(X_0\)  & 96.89 ± 0.91\sign & 33.93 ± 1.35\sign \\
\bottomrule
\end{tabular} \end{scriptsize}
\end{table}
To study the effect of the prediction target on the sample quality, the velocity (\(v\)) was exchanged for simple noise (\(\varepsilon\)\red{, similar to \cite{Cho2024SliceConsistent_MICCAI2024, chen2025multiview}}) or image (\(X_0\)\red{, similar to \cite{syndiff, arslan2024self}}) prediction (Tab.~\ref{tab:abl_sampl}). Both parametrization alternatives result in significantly reduced performance in all sampling methods. Note that \emph{regression} sampling is not possible for \(\varepsilon\)-prediction.

\subsubsection{Generalized gamma correction} \begin{table}
\centering
\caption{\red{
Comparison of applying linear, simple, or generalized gamma corrections after regression sampling YODA for BraTS. \\
\ \sign \ and \signA \  mark significant differences to the uncorrected images and the generalized gamma correction, respectively. 
}}
\label{tab:abl_gamma}
\begin{tabular}{lcc}
\toprule
correction method &  SSIM ($\%, \uparrow$) & PSNR (dB, $\uparrow$)   \\
\midrule
generalized gamma \eqref{eq:gamma}         &    \textbf{88.584 ± 5.513}\sign\ns &  25.774 ± 3.080\sign\ns \\
simple gamma (\(a,b=0\))       &  88.513 ± 5.570\sign\signA &  \textbf{25.779 ± 3.049}\sign\ns \\
linear (\(\gamma=0\))      &  88.580 ± 5.511\sign\signA &  25.764 ± 3.077\sign\ns \\
uncorrected (\(a,c,\gamma = 0\))   &     88.335 ± 5.619\signA\ns &   25.637 ± 3.108\signA\ns \\
\bottomrule
\end{tabular}
\end{table} {\setlength{\tabcolsep}{0.8pt} \renewcommand{\arraystretch}{0.5} \newcommand{\imgwidth}{1.65cm}
\begin{large}
\begin{figure}
\centering
\begin{tabular}{ccccc}
acquired \hspace{0.05mm} & uncorrected & general. $\Gamma$ & linear & $\Gamma$ \\
\includegraphics[width=\imgwidth]{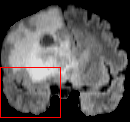} \hspace{0.05mm} & 
\includegraphics[width=\imgwidth]{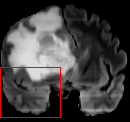} & \includegraphics[width=\imgwidth]{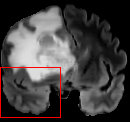} & \includegraphics[width=\imgwidth]{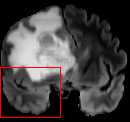} & \includegraphics[width=\imgwidth]{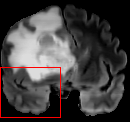} \\
\includegraphics[width=\imgwidth]{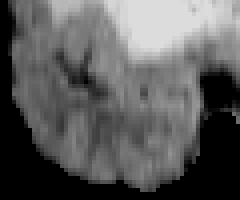} \hspace{0.05mm} & \includegraphics[width=\imgwidth]{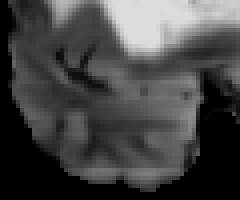} & \includegraphics[width=\imgwidth]{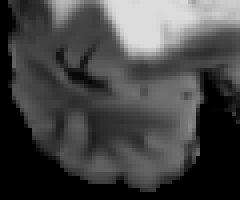} & \includegraphics[width=\imgwidth]{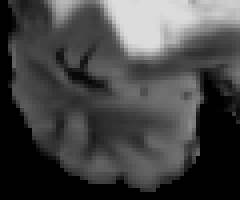} & \includegraphics[width=\imgwidth]{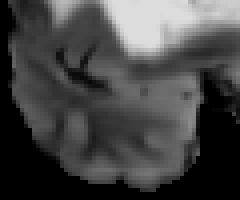} \\
\end{tabular}
    \caption{    
    \red{Left to right: 
    Acquired and predicted (regression, axial-only) in BraTS. Slicing artifacts were removed using a generalized Gamma (\(\Gamma\)), a linear (\(\gamma = 0\)), or simple \(\Gamma\) (\(a,b =0\)) correction \eqref{eq:gamma}.}
    } \label{fig:abl_gamma} 
\end{figure} \end{large} }
\red{To demonstrate the effect of the proposed generalized gamma correction \eqref{eq:gamma} on improving the 3D slice coherency in \emph{regression} sampling, we compared optimizing all parameters to simplifying to a linear (fixing \(\gamma=0\)) or an usual gamma correction (\(a=c=0\)). We present the results in Tab.~\ref{tab:abl_gamma} and Fig.~\ref{fig:abl_gamma}). \\
We found that both linear and simple gamma corrections improved the image quality (SSIM, PSNR) over the uncorrected baseline. Yet, the proposed generalized gamma including both a linear and non-linear component showed significantly higher SSIM than the linear and the simple gamma corrections. 
Conversely, the PSNR of the simple gamma correction was slightly but not significantly higher (\(p=0.11\)), which we attribute to potential additional shift in the intensity histograms. 
}

\section{Discussion}


In this work, we systematically assessed the presumed benefits of DMs for MIT.
While we confirm that the iterative refinement in DM sampling improves perceptual quality, we find that this implies the replication of inherent and undesired acquisition noise, 
\red{which degrades image fidelity and hinders \emph{all} evaluated downstream tasks. Notably, \emph{diffusion} sampling starts from the same initial prediction as in \emph{regression} sampling, meaning DMs already ``know'' the best solution but deliberately \emph{worsen} it through iterative refinement. This not only consumes vast computational resources but also degrades image quality (e.g., SSIM, PSNR) effectively ``re-noising'' clean predictions to produce more realistic, yet less informative outputs. } \\
We further show that perceptual realism can be trivially increased by adding noise to smooth images and that metrics like the FID favor realism over information fidelity. We observed the same for paired feature similarities (e.g., for cosine or \(L_2\) similarity) as supposedly more semantic measures \cite{blau2018perception_distortion_tradeoff}. This underscores a core limitation of such perceptual metrics: they reward the ability to mimic real (i.e.\ noisy images \red{\(X {\sim} \mathcal{X}\)}) rather than accurately reconstructed noise-free versions (\red{\(X' {\sim} \mathcal{X'}\)}). 
Our experiments confirm that methods like DDIM sampling or our ExpA gradually trade perceptual quality for reduced distortion. 
Thus, reported gains from fast DM sampling \cite{jiang2024fastddpm, syndiff, Xin2024Crossconditioned_MICCAI2024} or \red{ExpA-like} averages \cite{zhou2024cascaded, Cho2024SliceConsistent_MICCAI2024} \red{\cite{chen2025multiview, lyu2022conversion}} likely stem from noise suppression rather than truly improved image generation. In contrast, \emph{regression} sampling sacrifices perceptual quality to minimize distortion and maintain fidelity. 
Across all MRI tasks, smoother \emph{regression} images consistently outperformed \emph{diffusion}-generated images in downstream evaluations -- even when effects of domain shifts (\textit{SynthSeg}  \cite{billot2023synthseg}, internal RS WMH pipeline \cite{lohner2022relation}) can be ruled out.
Although clinicians may be more accustomed to noisy, realistic-looking images, noise suppression improves contrast and could actually enhance manual assessment and visual inspection. 


For brain MRI translation, ExpA-sampled DMs with \(N_\text{Ex}{=}10\) performed slightly worse than \emph{regression} sampling,
likely due to residual noise. \red{ However, we observed that increasing \(N_\text{Ex}\) leads to convergence: ExpA averaged \textit{diffusion} samples become -- qualitatively and quantitatively -- almost indistinguishable from \emph{regression} predictions. 
This convergence, however, demands thousands of network evaluations -- rendering it computationally impractical. Thus, at high \(N_\text{Ex} \gg 10\), DMs essentially recover their initial \emph{regression} output at an exorbitant cost and the expected value of \textit{diffusion} sampling corresponds to the initial \emph{regression}-like prediction. }
\red{
For the MRI $\rightarrow$ CT translation in the Gold Atlas dataset, \emph{diffusion} sampling marginally outperformed \emph{regression} sampling at $N_\text{Ex}=10$, possibly due to weaker conditioning in this task \cite{lyu2022conversion}. Still the small test set (two participants in \cite{lyu2022conversion} and four in our test) limits the strength and generalizability of this finding.} \\
\red{
We also explored alternative noise schedules or prediction targets, consistently confirming that ExpA drives DMs toward the \emph{regression} baseline. While our analysis is grounded in DDPMs \cite{ho2020denoising} the conclusions should extend to broader probabilistic DMs.
Future work should validate this for score-based \cite{song2020score} and flow-matching models \cite{lipmanflow}. 
} \\
\red{
Given how closely DMs' first denoising step approximates the target image, using source images as DB priors \cite{arslan2024self, Cho2024SliceConsistent_MICCAI2024} in strongly conditioned settings may be redundant or suboptimal. 
While some approaches split learning between semantic content and perceptual refinement via dual models 
\cite{zhou2024cascaded,chen2025multiview} 
this separation is implicitly unified in \textit{YODA}'s truncated diffusion strategy -- offering an efficient solution. }


In practical terms, using smooth \emph{regression} predictions achieved a comparable or better accuracy in downstream applications, except for glioma segmentation on BraTS. We attribute this exception to BraTS' known issues, such as poor image quality and imprecise image co-registration \cite{kong2021breaking}, which reduce information-richness in the source images and distort segmentation maps through misalignment. While we noted reduced quality metrics of the RS-trained \textit{YODA} on the external MBB data, this can at least be partially attributed to the distinctive FLAIR sequence in MBB. Notably, \textit{YODA} enables WMH segmentation without FLAIR, for the first time to our knowledge, when combined with established WMH tools \cite{shiva_wmh, lohner2022relation}.
\red{This opens up the possibility of retrospective lesion analysis in FLAIR-less datasets, such as the Human Connectome Project (HCP) \cite{glasser2016hcp}.} \\
\red{
Note that this setting can be seen as leveraging paired training data from the RS for a domain-adapting WMH segmentation, i.e.\ transferring FLAIR-based labels to studies without FLAIR acquisition.
However, training \textit{YODA} currently still requires paired data from source and target modalities.
To generalize further, future work could extend \textit{YODA} to unpaired MIT, possibly via leveraging cycle-consistent training \cite{zhu2017unpaired}. As adversarial losses are crucial to enforce domain realism \cite{zhu2017unpaired}, AmbientGANs \cite{bora2018ambientgan} may offer a promising path to disentangle noise replication and domain alignment.
} 

\red{
Finally, whereas DMs may offer value in weakly conditioned tasks like inpainting \cite{durrer2024denoising, pollak2025lit} or when used for data augmentation \cite{fernandez2022can, zhang2024diffboost}}, our results challenge the assumed superiority of DMs, and by extension GANs, in well-conditioned medical applications. Their popularity likely stems from natural image applications, misapplied to medical contexts where realism is often irrelevant, and information fidelity is paramount. This may be further amplified by innovation biases in method development \cite{isensee2024nnu_revisited}.


\subsection*{Conclusion} 
\red{Our findings highlight the domain differences between natural and medical image generation. We demonstrate the importance of accounting for noise replication and find that simple regression models outperform complex DMs and GANs in preserving medically relevant information. This suggests that, for MIT, fidelity rather than realism should guide model development and evaluation. We therefore recommend to always compare new methods to fair regression baselines and supplement the image quality assessment with relevant downstream tasks.}


\end{document}